\documentclass[a4paper]{article}
\usepackage{amsmath,amssymb,theorem,fullpage}
\usepackage[nosort]{cite}

\let\frac\undefined

\allowdisplaybreaks

\numberwithin{equation}{section}

\def\Maketitle{{\def\newpage{}\maketitle}}

\def\eq#1{\begin{equation}#1\end{equation}}
\long\def\subeq#1{\begin{subequations}\ignorespaces#1\end{subequations}}
\def\Split#1{\begin{split}#1\end{split}}
\def\Align#1{\begin{align}#1\end{align}}
\def\AlignS#1{\begin{align*}#1\end{align*}}
\def\Aligned#1{\begin{aligned}#1\end{aligned}}
\def\Gather#1{\begin{gather}#1\end{gather}}
\def\Gathered#1{\begin{gathered}#1\end{gathered}}
\def\Multline#1{\begin{multline}#1\end{multline}}

\def\pMatrix#1{\begin{pmatrix}#1\end{pmatrix}}
\def\Cases#1{\begin{cases}#1\end{cases}}
\def\d{\partial}
\def\bd{\bar\partial}

\def\Res{\mathop{\rm Res\,}\limits}
\def\const{\mathop{\rm const}\nolimits}
\def\spanspace{\mathop{\rm span}\nolimits}
\def\cA{{\cal A}}
\def\cB{{\cal B}}
\def\cC{{\cal C}}
\def\cF{{\cal F}}

\def\cI{{\cal I}}
\def\cL{{\cal L}}
\def\cO{{\cal O}}
\def\cP{{\cal P}}
\def\cQ{{\cal Q}}
\def\cS{{\cal S}}
\def\cT{{\cal T}}
\def\cV{{\cal V}}
\def\cZ{{\cal Z}}
\def\bfP{{\bf P}}
\def\bfQ{{\bf Q}}
\def\bfa{{\bf a}}
\def\ve{\varepsilon}
\def\sh{\mathop{\rm sh}\nolimits}
\def\ch{\mathop{\rm ch}\nolimits}
\def\th{\mathop{\rm th}\nolimits}

\def\tg{\mathop{\rm tg}\nolimits}

\def\lcolon{\mathopen{\,:\,}}
\def\rcolon{\mathclose{\,:\,}}
\def\C{{\mathbb{C}}}

\def\Z{{\mathbb{Z}}}
\def\llangle{\mathopen{\langle\!\langle}}
\def\rrangle{\mathclose{\rangle\!\rangle}}

\def\vR{{\check R}}

\def\vG{{\check G}}

\def\e{{\rm e}}
\def\i{{\rm i}}

\def\tf{{\tilde f}}
\def\tJ{{\tilde J}}

\def\tV{{\tilde V}}
\def\bT{{\bar T}}
\def\bz{{\bar z}}

\def\rad{{\rm rad}}
\def\vac{{\rm vac}}

\def\Eta{{\rm H}}

\def\vpj{{\vphantom{j}}}

\makeatletter

\def\section{\@startsection{section}{1}{\z@}%
                                   {-3.5ex \@plus -1ex \@minus -.2ex}%
                                   {2.3ex \@plus.2ex}%
                                   {\normalfont\normalsize\bfseries}}
\def\subsection{\@startsection{subsection}{2}{\z@}%
                                     {-3.25ex\@plus -1ex \@minus -.2ex}%
                                     {1.5ex \@plus .2ex}%
                                     {\normalfont\normalsize\bfseries\itshape}}
\def\@seccntformat#1{\csname the#1\endcsname.~~}
\long\def\@makecaption#1#2{%
  \vskip\abovecaptionskip
  \sbox\@tempboxa{\small#1. #2}%
  \ifdim \wd\@tempboxa >0.9\hsize
  {\leftskip=0.05\hsize\rightskip=0.05\hsize\relax\small
    #1. #2\par}
  \else
    \global \@minipagefalse
    \hb@xt@\hsize{\hfil\box\@tempboxa\hfil}%
  \fi
  \vskip\belowcaptionskip}
\def\Appendix{\appendix
  \def\@seccntformat##1{Appendix~\csname the##1\endcsname.~~}}

\let\over\@@over
\let\atop\@@atop
\let\above\@@above
\let\overwithdelims\@@overwithdelims
\let\atopwithdelims\@@atopwithdelims
\let\abovewithdelims\@@abovewithdelims

\makeatother

\newtheorem{theorem}{Theorem}
\newtheorem{conjecture}{Conjecture}
\newtheorem{corollary}{Corollary}

\begin{document}


\title{Form factors of descendant operators:\\ Free field construction and reflection relations}
\author{Boris Feigin$^{1,2}$ and Michael Lashkevich$^1$\\[\medskipamount]
$^1$~\parbox[t]{0.9\textwidth}{\normalsize\it\raggedright
Landau Institute for Theoretical Physics,
142432 Chernogolovka of Moscow Region, Russia}\\
$^2$~\parbox[t]{0.9\textwidth}{\normalsize\it\raggedright
Independent University of Moscow, 11 Bolshoy Vlasyevsky pereulok, 119002 Moscow, Russia}}

\date{}

\Maketitle
\vskip -\medskipamount
\rightline{\it\large Dedicated to the memory of Alexey Zamolodchikov}
\bigskip

\begin{abstract}
The free field representation for form factors in the sinh-Gordon model and the sine-Gordon model in the breather sector is modified to describe the form factors of descendant operators, which are obtained from the exponential ones, $\e^{\i\alpha\varphi}$, by means of the action of the Heisenberg algebra associated to the field~$\varphi(x)$. As a check of the validity of the construction we count the numbers of operators defined by the form factors at each level in each chiral sector. Another check is related to the so called reflection relations, which identify in the breather sector the descendants of the exponential fields $\e^{\i\alpha\varphi}$ and $\e^{\i(2\alpha_0-\alpha)\varphi}$ for generic values of~$\alpha$. We prove the operators defined by the obtained families of form factors to satisfy such reflection relations. A generalization of the construction for form factors to the kink sector is also proposed.
\end{abstract}

\section{Introduction}

Exact calculation of form factors of local and quasilocal operators in two-dimensional relativistic quantum field theory is known to be reduced to solving a set of difference equation for analytic functions~\cite{Karowski:1978vz,Smirnov:1984sx,Smirnovbook} called also Karowski--Weisz--Smirnov form factor axioms. One of the techniques for solving these equations is the free field representation proposed by Lukyanov~\cite{Lukyanov:1993pn}. It was shown that this representation makes it possible to calculate form factors of the exponential fields $\e^{\i\alpha\varphi}$ in the sine/sinh-Gordon model~\cite{Lukyanov:1997bp}. But the family of exponential operators is far from exhausting the full set of operators in the theory, which contains also the descendant operators obtained from the exponential ones by means of the action of the Heisenberg algebra associated with the field~$\varphi(x)$. Here we propose a construction of the form factors of descendant operators in the breather sector of the sine-Gordon theory and for the sinh-Gordon theory.

We start from the proposal of Babujian and Karowski~\cite{Babujian:1999ht,Babujian:2002fi}, who expressed the form factors of descendant operators in terms of sequences of some auxiliary functions. These sequences must satisfy some conditions to provide form factors of local operators. The main distinction of our approach is that we impose much more restrictive conditions to these sequences of functions, which makes it possible to substantiate the existence of a one-to-one correspondence between operators and sequences of functions.

Besides, we propose an interpretation of these solutions in terms of an auxiliary commutative algebra and show that at a generic value of $\alpha$ the dimensions of the level subspaces coincide with those for the Fock modules of the Heisenberg algebra. We go further and, by means of some auxiliary bosonization procedure, prove the existence of a reflection property, which relates breather form factors of descendants of the fields $\e^{\i\alpha\varphi}$ and $\e^{\i(2\alpha_0-\alpha)\varphi}$ (with the value of $\alpha_0$ known from the conformal field theory). Earlier it was conjectured that such relations, well known in the Liouville field theory~\cite{Zamolodchikov:1995aa}, are valid for the operators in the sinh-Gordon theory~\cite{Fateev:1997nn,Fateev:1998xb}. One may expect that they are valid for the sine-Gordon theory in the sector corresponding to the perturbed minimal model~\cite{Smirnov:1989hh,Smirnov:1990vm}. Surely, this sector includes the breathers. Notice, that our approach has much in common with that of~\cite{Babelon:1996sk}, though we concentrate our attention on the breather sector at generic values of the coupling constants and field parameters.

\section{Operator contents of the sine/sinh-Gordon model}

Consider the sine-Gordon model
\eq{
S_{\textrm{SG}}[\varphi]=\int d^2x\,\left({(\d_\mu\varphi)^2\over8\pi}
+\mu\cos\beta\varphi\right).
\label{SGdef}
}
We shall also use the notation
$$
\beta^2=2{p\over p+1}\le2,
\qquad
\alpha_0={1\over\beta}-{\beta\over2}={1\over\sqrt{2p(p+1)}}.
$$
The spectrum of the sine-Gordon model consists of a kink--antikink (or a soliton--antisoliton) pair of some mass $M$, which can be expressed in terms of to the parameter $\mu$~\cite{Zamolodchikov:1995xk}, and a series of breathers that is nonempty for $\beta^2\le1$. The masses of the breathers are given by the formula
\eq{
m_n=2M\sin{\pi pn\over2},
\qquad
n=1,2,\ldots,
\qquad
pn\le1.
\label{m_n-def}
}
Besides, the higher breathers may be considered as bound states of the first breathers of the mass $m=m_1$. It means that the form factors with respect to the states consisting any $n$-breathers can be expressed in terms of the form factors with respect to the states only containing the $1$-breathers. That is why we restrict our consideration to this $1$-breather, which will be just called breather hereafter.

We can also consider the sinh-Gordon model
\eq{
S_{\textrm{ShG}}[\varphi]=\int d^2x\,\left({(\d_\mu\varphi)^2\over8\pi}
+\mu\ch\hat\beta\varphi\right).
\label{ShGdef}
}
The spectrum of the model consists of the only kind of particles, which can be considered as an `analytic continuation' of the $1$-breather in the following sense. The expressions for the form factors of every local operator with respect to these particle coincide with those with respect to the states consisting of the $1$-breathers after the substitution $\beta\to-\i\hat\beta$. Hence, the sinh-Gordon model corresponds to the region~$-1<p<0$.

The $S$ matrix of two breathers is
\eq{
S(\theta)={\th{\theta+\i\pi p\over2}\over\th{\theta-\i\pi p\over2}}.
\label{Smatrix}
}

Consider the operator contents of the models. Let us start with the exponential operators
\eq{
V_{(\alpha)}(x)=\e^{\i\alpha\varphi(x)}.
\label{expdef}
}
Below it will be more convenient to use another parameter
\eq{
a={\alpha-\alpha_0\over2\alpha_0}.
\label{a-alpha-def}
}
We shall always assume that the parameters $a$ and $\alpha$ are related according to~(\ref{a-alpha-def}). Since we want to use both letters as subscripts, we shall always use $\alpha$ there in parentheses and $a$ without them, e.~g.
$$
V_{(\alpha)}(x)\equiv V_a(x).
$$

The exponential operators do not exhaust the operator contents of the theory. We have to define the descendant operators. First of all, recall that at small enough distances the field theories (\ref{SGdef}) and (\ref{ShGdef}) behave like a free boson theory. Take any point in the Euclidean plane, e.~g.\ $x=0$, and consider the radial quantization picture around this point. Let
$$
z=x^1-x^0=x^1+\i x^2,
\qquad
\bz=x^1+x^0=x^1-\i x^2,
\qquad
\d={\d\over\d z},
\qquad
\bd={\d\over\d\bz}.
$$
The radial quantization means that we consider radial coordinates $\sigma$, $\tau$ such that
$$
z=\e^{\tau+\i\sigma}
$$
and consider $\tau$ as an imaginary time and $\sigma$ as a space dimension. There is a one-to-one correspondence between states $|\cO\rangle_\rad$ in this picture and local operators $\cO(x)$ put to the point $x=0$. This correspondence survives the perturbation for nearly all fields except some particular (`resonant') operators.

The free field $\varphi(x)$ can be expanded in this picture as
\eq{
\varphi(x)=\bfQ-\i\bfP\log z\bz+\sum_{n\ne0}{\bfa_n\over\i n}z^{-n}
+\sum_{n\ne0}{\bar\bfa_n\over\i n}\bar z^{-n}.
\label{varphi-expansion}
}
The operators $\bfQ$, $\bfP$, $\bfa_n$, $\bar\bfa_n$ form a Heisenberg algebra with the only nonzero commutation relations
\eq{
[\bfP,\bfQ]=-\i,
\qquad
[\bfa_m,\bfa_n]=m\delta_{m+n,0},
\qquad
[\bar\bfa_m,\bar\bfa_n]=m\delta_{m+n,0}.
\label{PQaa-alg}
}
The states $|\alpha\rangle_\rad$ defined as
\eq{
\bfa_n|\alpha\rangle_\rad=\bar\bfa_n|\alpha\rangle_\rad=0\quad(n>0),
\qquad
\bfP|\alpha\rangle_\rad=\alpha|\alpha\rangle_\rad,
\qquad
|\alpha\rangle_\rad=\e^{\i\alpha\bfQ}|0\rangle_\rad
\label{alpharad}
}
correspond (up to a constant factor) to the operators~$V_{(\alpha)}(x)$. The descendants form a Fock module of the algebra (\ref{PQaa-alg}) with the highest vector~$|\alpha\rangle_\rad$. We may choose the basis%
\footnote{The states obtained by the action of $\bfQ^m$ (corresponding to the operators containing $\varphi^m$) can be obtained as the $m$th derivatives in $\alpha$ and thus are obtained trivially.}
\eq{
\bfa_{-k_1}\ldots\bfa_{-k_s}\bar\bfa_{-l_1}\ldots\bar\bfa_{-l_t}|\alpha\rangle_\rad
\quad\leftrightarrow\quad{(-\i)^{s+t}\over\prod(k_i-1)!\prod(l_j-1)!}\,
\d^{k_1}\varphi\ldots\d^{k_s}\varphi\bd^{l_1}\varphi\ldots\bd^{l_t}\varphi
\e^{\i\alpha\varphi}
\label{descendants-def}
}
with $0<k_1\le\ldots\le k_s$, $0<l_1\le\ldots\le l_t$. The pair of integers $(n,\bar n)$, where $n=\sum k_i$, $\bar n=\sum l_i$, is called the level of the element. The integers $n$ and $\bar n$ are called chiral and antichiral level correspondingly. The descendants only generated by the elements $\bfa_{-k}$ are usually called chiral descendants, while those only generated by the elements $\bar\bfa_{-l}$ are called antichiral ones. Let $\cF$ be the Fock submodule of chiral descendants. The submodule of antichiral descendants will be referred to as $\bar\cF$. Evidently, the submodules $\cF$ and $\bar\cF$ are isomorphic. The Fock module spanned on all the vectors (\ref{descendants-def}) is the tensor product $\cF\otimes\bar\cF\simeq\cF\otimes\cF$. The module $\cF$ admits a natural gradation into the subspaces $\cF_n$ by the chiral level~$n$:
$$
\cF=\bigoplus^\infty_{n=0}\cF_n,
\qquad
\cF_n=\spanspace\Bigl\{\bfa_{-k_1}\ldots\bfa_{-k_s}|\alpha\rangle_\rad\Big|
s\in\Z_{\ge0},\sum^s_{i=1}k_i=n\Bigr\}.
$$
The generating function of dimensions of these subspaces (the character) is given by
\eq{
\chi(q)\equiv\sum^\infty_{n=0}q^n\dim\cF_n=\prod^\infty_{k=1}{1\over1-q^k}.
\label{chidef}
}

\section{Form factors from free field representation}

\subsection{Form factors of exponential operators}

First let us describe the form factors of exponential operators. Let $|\vac\rangle$ be the vacuum and $|\theta_1,\ldots,\theta_N\rangle$ be the eigenstate of the Hamiltonian corresponding to $N$ breathers with rapidities $\theta_1<\ldots<\theta_N$. The form factors of the exponential can be written as
\eq{
\langle\theta_{k+1},\ldots,\theta_N|V_a(0)|\theta_1,\ldots,\theta_k\rangle
=G_af_a(\theta_1,\ldots,\theta_k,
\theta_{k+1}+\i\pi,\ldots,\theta_N+\i\pi).
\label{fa-def}
}
Here $G_a$ is the vacuum expectation value, which is known exactly~\cite{Lukyanov:1996jj}:
\Align{
G_a
&\equiv\langle V_a(0)\rangle=m^{\alpha^2}\vG_a
\notag
\\*
&=\left(m{\Gamma\left(1+p\over2\right)\Gamma\left(2-p\over2\right)
\over4\sqrt\pi}\right)^{\alpha^2}
\exp\int^\infty_0{dt\over t}\,
\left({\sh{t\over2}\sh^2(a+{1\over2})t\over\sh t\sh{pt\over2}\sh{(p+1)t\over2}}
-{(2a+1)^2\over2p(p+1)}e^{-(p+1)t}\right).
\label{VaVEV}
}
Using the free field representation~\cite{Lukyanov:1993pn} the analytic functions $f_a(\theta_1,\ldots,\theta_N)$ are expressed in terms of trace functions of vertex operators~\cite{Lukyanov:1997bp}. Omitting the details let us write the answer in the form
\eq{
f_a(\theta_1,\ldots,\theta_N)
=\llangle T(\theta_N)\ldots T(\theta_1)\rrangle_a.
\label{fa-trace}
}
Here $T(\theta)$ is a generator of the degenerate deformed Virasoro algebra~\cite{Lukyanov:1995gs} and $\llangle\dots\rrangle_a$ is a trace function with the property
$$
\llangle XT(\theta)\rrangle_a=\llangle T(\theta+2\pi\i)X\rrangle_a
\qquad(\forall X).
$$
The generator $T(\theta)$ can be written in the form
\eq{
T(\theta)
=\i\lambda'\left(\e^{-\i\pi\hat a}\Lambda_+(\theta)
+\e^{\i\pi\hat a}\Lambda_-(\theta)\right)
\label{Tdef}
}
with the constant factor
$$
\lambda'=\left(1\over2\sin{\pi p\over2}\right)^{1/2}
\exp\left(-\int^{\pi p}_0{dt\over2\pi}\,{t\over\sin t}\right).
$$
The element $\hat a$ is central so that
\eq{
\llangle F(\hat a)\rrangle_a=\llangle F(a)\rrangle
\qquad
(\forall F).
\label{alphahatdef}
}
The pair trace functions of the vertex operators $\Lambda_\pm$ are given by
\eq{
\Gathered{
\Aligned{
\llangle\Lambda_\pm(\theta_2)\Lambda_\pm(\theta_1)\rrangle
&=R(\theta_1-\theta_2),
\\
\llangle\Lambda_\pm(\theta_2)\Lambda_\mp(\theta_1)\rrangle
&=R^{-1}(\theta_1-\theta_2\mp\i\pi)
=R(\theta_1-\theta_2)
{\sh(\theta_1-\theta_2)\pm\i\sin\pi p\over\sh(\theta_1-\theta_2)},
}
\\
R(\theta)=\exp\left(
4\int^\infty_0{dt\over t}\,
{\sh{\pi t\over2}\sh{\pi pt\over2}\sh{\pi(p+1)t\over2}
\over\sh^2\pi t}\ch(\pi-\i\theta)t
\right),
}\label{LambdaLambda}
}
while the general trace functions in the right hand side of (\ref{alphahatdef}) factorize into pair trace functions of the vertex operators~$\Lambda_\ve(\theta)$:
\eq{
\llangle\Lambda_{\ve_N}(\theta_N)\ldots\Lambda_{\ve_1}(\theta_1)\rrangle
=\prod_{1\le i<j\le N}\llangle\Lambda_{\ve_j}(\theta_j)
\Lambda_{\ve_i}(\theta_i)\rrangle.
\label{llrrdef}
}

More explicitly, the functions $f_a$ may be written as
\eq{
f_a(\theta_1,\ldots,\theta_N)
=(\i\lambda')^N\sum_{\ve_1,\ldots,\ve_N=\pm}
\prod^N_{i=1}\e^{-\i\pi a\ve_i}\cdot
\llangle\Lambda_{\ve_N}(\theta_N)\ldots\Lambda_{\ve_1}(\theta_1)\rrangle.
\label{fdef}
}

\subsection{Generalization to descendant operators}

The question is how to generalize the expression (\ref{fdef}) to the descendant operators. Babujian and Karowski~\cite{Babujian:1999ht,Babujian:2002fi} proposed the following generalization:%
\eq{
f^P_a(\theta_1,\ldots,\theta_N)
=(\i\lambda')^N\sum_{\ve_1,\ldots,\ve_N=\pm}
P_{\ve_1\ldots\ve_N}(\e^{\theta_1},\ldots,\e^{\theta_N})
\prod^N_{i=1}\e^{-\i\pi a\ve_i}\cdot
\llangle\Lambda_{\ve_N}(\theta_N)\ldots\Lambda_{\ve_1}(\theta_1)\rrangle.
\label{fPdef}
}
The paper~\cite{Babujian:2002fi} is based on some particular choice of the $P$ functions for particular fields, so that the analyticity of the results in the parameter $\alpha$ is hidden there. We propose here a more systematic way to count and study descendants, so that the analytic properties were always apparent. Our strategy is the opposite: if Babujian and Karowski allow the $P$ functions to be as arbitrary as possible, we, on the contrary, impose as many restrictions as possible aiming to establish a bijection between consequences of the $P$ functions and operators, at least for generic values of~$p$ and~$a$. Moreover, any exceptional cases will always be considered here as limits from the generic point.

The functions $P_{\ve_1\ldots\ve_N}(x_1,\ldots,x_N)$ are supposed to be entire functions of the variables $x_1,\ldots,x_N$. The functions $f^P_\alpha$ satisfy the form factor axioms subject to the following restrictions:
\Gather{
P_{\ldots\ve\ve'\ldots}(\ldots,x,x',\ldots)
=P_{\ldots\ve'\ve\ldots}(\ldots,x',x,\ldots),
\label{Psym}
\\
P_{-+\ve_1\ldots\ve_N}(-x,x,x_1,\ldots,x_N)
=P_{\ve_1\ldots\ve_N}(x_1,\ldots,x_N).
\label{Pkin}
}
Though this condition is more restrictive than that of~\cite{Babujian:2002fi}, it is yet too weak. In particular, such general form admits a solution of the form $\prod\e^{-\i\pi\,\delta a\,\ve_i}$, which is equivalent to the substitution $a\to a+\delta a$. It means that we have to impose some extra restrictions on the functions~$P$. To formulate these new restrictions, let us reformulate the condition (\ref{Psym}),~(\ref{Pkin}).

Due to the symmetry property (\ref{Psym}) these functions can be uniquely expressed in terms of the functions
\eq{
P_{N,k}(x_1,\ldots,x_{N/2+k}|y_1,\ldots,y_{N/2-k})
=P_{\underbrace{-\ldots-}_{N/2+k}\underbrace{+\ldots+}_{N/2-k}}
(x_1,\ldots,x_{N/2+k},y_1,\ldots,y_{N/2-k}).
\label{Pkdef}
}
Here $k$ is an integer (a half integer) for even (odd) $N$. These function are symmetric with respect to the variables $X=(x_1,\ldots,x_{N/2+k})$ and with respect to the variables $Y=(y_1,\ldots,y_{N/2-k})$ separately. We also shall write them as $P_{N,k}(X|Y)$. We shall use the notation like this everywhere, if the dimensions of the arrays are clear from the context. We shall also use the notation like $X^n=(x_1^n,\ldots,x^n_{N/2+k})$, $\lambda X=(\lambda x_1,\ldots,\lambda x_N)$.

The kinematic pole condition takes the form
\eq{
P_{N+2,k}(X,-x|x,Y)=P_{N,k}(X|Y).
\label{Pk-kin}
}

First, let us restrict the class of functions $P_{N,k}$ with the Laurent polynomials in the variables. Each such Laurent polynomial can be rewritten as a sum
\eq{
P_{N,k}(X|Y)=\sum_AP_{N,k}^A(X|Y)\bar P_{N,k}^A(X^{-1}|Y^{-1}),
\label{Psumform}
}
where $P_{N,k}^A(X|Y)$, $\bar P_{N,k}^A(X|Y)$ are some homogeneous polynomials symmetric with respect to the variables $x_i$ and $y_j$ separately.

Let $p(\xi_1,\ldots,\xi_n|\eta_1,\ldots,\eta_n)$ be a polynomial of the degree $n$ in the following sense. Set the degree of the variables $\xi_i$ and $\eta_i$ equal to~$i$. The degree of each monomial in these variables is the sum of the degrees of each variable. Then define a family of the homogeneous polynomials $P^{[p]}_{N,k}(X,Y)$ of the order $n$ for any $N$ and $k$ as follows:
\eq{
P^{[p]}_{N,k}(X|Y)=p(S_1(X),\ldots,S_n(X)|S_1(Y),\ldots,S_n(Y)),
\label{Pk-p}
}
where $S_r(X)$ are power sums of the order $r$:
\eq{
S_r(X)=\sum_i x_i^r
\label{Skdef}
}
It is known that, due to the Newton--Girard identities~\cite{NGids}, any symmetric polynomial can be written as a polynomial of power sums. Here it will be more convenient to use the power sums rather than the more usual elementary symmetric polynomials.

Our second restriction is that each of the polynomials $P^A_{N,k}$, $\bar P^A_{N,k}$ has the form (\ref{Pk-p}) with an appropriate $n$ and $p$. In other words, each of these polynomials can be expressed in terms of elementary symmetric polynomials in the same form independently of the values of~$N$ and~$k$. In particular, for the constant solutions of (\ref{Pk-kin}) this condition means that the all $P_{N,k}$ coincide for every $N$ and~$k$, excluding the products that shift the value of~$a$.

Let $\cP_n$ be the space of the order $n$ polynomials $p(\xi_1,\ldots,\xi_n|\eta_1,\ldots,\eta_n)$ such that the polynomials $P^{[p]}_{N,k}$ defined in~(\ref{Pk-p}) satisfy the equation~(\ref{Pk-kin}) for any $N$ and~$k$.

\begin{theorem}\label{th-char}
The generating functions of the dimensions of the spaces $\cP_n$ coincides with the character of the Fock module~$\cF$:
$$
\sum^\infty_{n=0}q^n\dim\cP_n=\chi(q).
$$
\end{theorem}

The Theorem~\ref{th-char} means that the dimension of the space $\cP_n$ coincides with the dimension of the level $n$ subspace $\cF_n$ of the Fock module~$\cF$:
$$
\dim\cP_n=\dim\cF_n.
$$

\begin{conjecture}\label{conjecture1}
There is a one-to-one correspondence between the level $n$ subspace $\cF_n$ of the Fock module $\cF$ and the space $\cP_n$. Each element $p$ of $\cP_n$ defines a level $n$ chiral Fock descendant of the operator $V_a(x)$ by its form factors according to (\ref{fPdef}), (\ref{Pkdef}),~(\ref{Pk-p}).
\end{conjecture}

\subsection{Proof of the Theorem~\ref{th-char}}
\label{th1-proof}

There is a constructive way to get all solutions. Let $\cA$ be the Abelian algebra generated by the elements $c_{-1},c_{-2},\ldots$. Consider two currents:
\eq{
\Aligned{
a(z)
&=\e^{\sum^\infty_{m=1}c_{-m}z^m},
\\
b(z)
&=\e^{-\sum^\infty_{m=1}c_{-m}(-z)^m},
}\label{abcurrents}
}
such that
\eq{
a(z)b(-z)=1.
\label{abprod}
}
Define the inner product in the algebra $\cA$:
$$
\left(\prod^\infty_{m=1}c_{-m}^{k_m},
\prod^\infty_{m=1}c_{-m}^{l_m}\right)
=\prod^\infty_{m=1}k_m!\>\delta_{k_ml_m}.
$$
For any element $h\in\cA$ define a function
\eq{
P_{N,k}^h(X|Y)=\left(a(x_1)\ldots a(x_{N/2+k})
b(y_1)\ldots b(y_{N/2-k}),h\right).
\label{Pkhdef}
}
If $h$ is a basic element of the order $n$,
\eq{
h=c_{-1}^{k_1}\ldots c_{-s}^{k_s},
\qquad \sum^s_{m=1}mk_m=n,
\label{basicel}
}
the functions $P_{N,k}^h(X|Y)$ are a polynomials with the necessary properties. The integer $n$ will be called the level of the element. The subspace spanned on level $n$ elements will be denoted as~$\cA_n$, $\dim\cA_n=\dim\cP_n$.

The basic elements (\ref{basicel}) are in a one-to-one correspondence with the Fock vectors, given by the map $c_{-m}\mapsto a_{-m}$ to the corresponding generators of the Fock algebra: $[a_k,a_l]=k\delta_{k+l,0}$.

The basic polynomials (\ref{Pkhdef}) corresponding to the elements (\ref{basicel}) can easily be written explicitly. Indeed,
$$
a(x_1)\ldots a(x_{N/2+k})b(y_1)\ldots b(y_{N/2-k})
=\prod^\infty_{m=1}\e^{\left(\sum^{N/2+k}_{i=1}x_i^m
-(-1)^m\sum^{N/2-k}_{i=1}y_i^m\right)c_{-m}}.
$$
Therefore,
$$
\left(a(x_1)\ldots a(x_{N/2+k})b(y_1)\ldots b(y_{N/2-k}),h\right)
=\prod^s_{m=1}\left(\e^{\left(\sum^{N/2+k}_{i=1}x_i^m
-(-1)^m\sum^{N/2-k}_{i=1}y_i^m\right)c_{-m}},c_{-m}^{k_m}\right).
$$
As a result we have
\Align{
P_{N,k}^h(X|Y)
&=\prod^s_{m=1}\left(\sum^{N/2+k}_{i=1}x_i^m
-(-1)^m\sum^{N/2-k}_{i=1}y_i^m\right)^{k_m}
\notag
\\
&=\prod^s_{m=1}(S_m(X)-(-1)^mS_m(Y))^{k_m}
\qquad
\text{for $h=c_{-1}^{k_1}\ldots c_{-s}^{k_s}$}.
\label{Pkh-explicit}
}
This proves that the sets of polynomials can be written in the form~(\ref{Pk-p}).

Besides, it proves linear independence of these sets of polynomials. Indeed, the functions $z_m=\sum^{N/2+k}_{i=1}x_i^m-(-1)^m\sum^{N/2-k}_{i=1}y_i^m$ are functionally independent for large enough $N$. Hence, linear independence of the sets of polynomials reduces to the evident linear independence of the monomials $z_1^{k_1}\ldots z_s^{k_s}$.

\subsection{Algebraic representation of form factors}

The above construction makes it possible to describe form factors in purely algebraic terms.

Consider two copies of the algebra $\cA$, which will be denoted as $\cA$ and $\bar\cA$, generated by the generators $\{c_{-n}\}$ and $\{\bar c_{-n}\}$ correspondingly. Define a natural homomorphism $\cA\to\bar\cA$: for any $h\in\cA$ we define $\bar h\in\bar\cA$ according to the rule $c_{-n}\mapsto\bar c_{-n}$. An element $g=h\bar h'$, $h\in\cA_n$, $h'\in\cA_{\bar n}$ will be referred to as a level $(n,\bar n)$ element. Let
\eq{
\cT(\theta)=
\i\lambda'\left(
\e^{-\i\pi\hat a}b(\e^\theta)\bar a(\e^{-\theta})
\Lambda_+(\theta)
+\e^{\i\pi\hat a}a(\e^\theta)\bar b(\e^{-\theta})
\Lambda_-(\theta)
\right).
\label{cTdef}
}
Let $g\in\cA\otimes\bar\cA$ be an arbitrary element. Then
\eq{
f^g_a(\theta_1,\ldots,\theta_N)
=\left(\llangle\cT(\theta_N)\ldots\cT(\theta_1)\rrangle_a,
g\right)
\label{fhdef}
}
be a form factor of an operator from the Fock space $(\cF\otimes\bar\cF)V_a(x)$. We shall denote the fields corresponding to the form factors $G_af^g_a(\theta_1,\ldots,\theta_N)$ as $V^g_a(x)$.%
\footnote{We preserve the factor $G_a$ here for the consistency of the notation only: $V^1_a(x)=V_a(x)$.}

The expression (\ref{fhdef}) is the `most algebraic' representation for the form factor. Let us write now the most explicit expression. First, define the functions
$$
P^g_{N,k}(X|Y)=(a(x_1)\bar b(x_1^{-1})\ldots a(x_{N/2+k})\bar b(x_{N/2+k}^{-1})
b(y_1)\bar a(y_1^{-1})\ldots b(y_{N/2-k})\bar a(y_{N/2-k}^{-1}),g)
$$
corresponding to an element $g\in\cA\otimes\bar\cA$. They are uniquely determined by the relations
\subeq{\Align{
P^{k_1g_1+k_2g_2}(X|Y)
&=k_1P^{g_1}(X|Y)+k_2P^{g_2}(X|Y),
\label{Ph1ph2}
\\*
P^{g_1g_2}(X|Y)
&=P^{g_1}(X|Y)P^{g_2}(X|Y)
\qquad\forall k_1,k_2\in\C,\ g_1,g_2\in\cA\otimes\bar\cA,
\label{Ph1th2}
\\
P^{c_{-n}}(X|Y)
&=S_n(X)-(-1)^nS_n(Y),
\label{Pcn}
\\*
P^{\bar c_{-n}}(X|Y)
&=S_{-n}(Y)-(-1)^nS_{-n}(X).
\label{Pcbarn}
}\label{Phdef}}
The subscripts $N$ and $k$ are omitted here. From (\ref{fPdef}) and (\ref{LambdaLambda}) we obtain
\eq{
f^g_a(\theta_1,\ldots,\theta_N)
=(\i\lambda')^N\prod^N_{i<j}R(\theta_i-\theta_j)
\cdot J^g_{N,a}
(\e^{\theta_1},\ldots,\e^{\theta_N}),
\label{fJ}
}
where
\eq{
J^g_{N,a}(x_1,\ldots,x_N)
=\sum_{I_++I_-=I}\e^{\i\pi a(\#I_--\#I_+)}P^g(X_-|X_+)
\prod_{i\in I_-\atop j\in I_+}f\left(x_i\over x_j\right)
\label{Jadef}
}
with
\eq{
f(x)={(x+\omega)(x-\omega^{-1})
\over x^2-1}=1+{\omega-\omega^{-1}\over x-x^{-1}},
\qquad
\omega=\e^{\i\pi p}.
\label{fxdef}
}
Here $I=\{1,\ldots,N\}$ and the sum is taken over all decompositions of $I$ into two subsets $I_+$ and $I_-$ ($I_+\cup I_-=I$, $I_+\cap I_-=\emptyset$). Besides,
$$
X_\pm=\{x_i|i\in I_\pm\}.
$$
The functions $J^g_{N,a}$ are symmetric in the variables $x_1,\ldots,x_N$. As we show below these functions possess some pleasant properties, and they are what the rest of our story is about.

The natural question is if two different elements $g_1$ and $g_2$ produce form factors of different operators according to Eq.~(\ref{fhdef}). We shall answer this question positively in the next subsection.

\subsection{Cluster property, holomorphic factorization and bijection property}

Let $g=h\bar h'$, where $h,h'\in\cA$. Let us calculate the asymptotics
$$
f^g_a(\theta_1,\ldots,\theta_l,
\theta_{l+1}+\Lambda,\ldots,\theta_N+\Lambda)
\quad\text{as $\Lambda\to+\infty$.}
$$
Take into account that $f(x\e^{\pm\Lambda})\to1$, $R(\theta\pm\Lambda)\to1$. Besides,
$$
P^h(X,X'\e^{-\Lambda}|Y,Y'\e^{-\Lambda})\to P^h(X|Y),
\qquad
P^{\bar h'}(X\e^\Lambda,X'|Y\e^\Lambda,Y')\to P^{\bar h'}(X'|Y').
$$
Therefore
$$
P^{h\bar h'}(X\e^\Lambda,X'|Y\e^\Lambda,Y')
\simeq P^h(X\e^\Lambda|Y\e^\Lambda)P^{\bar h'}(X'|Y')
$$
and
$$
J^{h\bar h'}_a(X\e^\Lambda,X')
\simeq J^{h}_a(X\e^\Lambda)
J^{\bar h'}_a(X').
$$
Thus we immediately get the following \textit{cluster factorization property}:
\eq{
f^{h\bar h'}_a(\theta_1,\ldots,\theta_l,
\theta_{l+1}+\Lambda,\ldots,\theta_N+\Lambda)
\simeq f^{h}_a(\theta_{l+1}+\Lambda,\ldots,\theta_N+\Lambda)
f^{\bar h'}_a(\theta_1,\ldots,\theta_l)
\
\text{as $\Lambda\to+\infty$}.
\label{cluster}
}
Comparing with the result of~\cite{Delfino:2005wi} we conclude that the form factors $f^h_a$, $h\in\cA_n$ correspond to level $n$ chiral descendants, while the form factors $f^{\bar h}_a$ correspond to the level $n$ antichiral ones.

Nevertheless, we can say nothing definite about the operator corresponding to an arbitrary element of the form $h\bar h'$, $h\in\cA_n$, $h\in\bar\cA_{\bar n}$, except that it is a linear combination of descendents of levels $(m,\bar m)$ such that $0\le m\le n$, $0\le\bar m\le\bar n$.

The expression~(\ref{fhdef}) defines a map $\Phi_a$ from the algebra $\cA\otimes\bar\cA$ to the space of consequences of analytic functions of $0,1,2,\ldots$ variables. Let $\cB_a$ be the image of the map~$\Phi_a$. We shall denote the image of the element $g$ as $f^g_a$ without arguments.

\begin{theorem}\label{theorem-bijection}
The map $\Phi_a:\cA\otimes\bar\cA\to\cB_a$ is a bijection for generic values of the parameter~$a$.
\end{theorem}

Let $g_1=h_1\bar h'_1$, $g_2=h_2\bar h'_2$. Due to the cluster factorization property two consequences $f^{g_1}_a$ and $f^{g_2}_a$ can coincide only if $f^{h_1}_a=f^{h_2}_a$ and $f^{\bar h'_1}_a=f^{\bar h'_2}_a$. Therefore it is enough to prove the theorem for the elements of the subalgebras $\cA\otimes1$ and $1\otimes\bar\cA$ separately. Consider e.~g.\ the first subalgebra.

First consider the map $\Phi_a$ in the limit $a\to-\i\infty$. The second term in (\ref{cTdef}) only survives this limit. Hence,
$$
\left.\e^{-\i\pi Na}J^h_a(x_1,\ldots,x_N)\right|_{a\to-\i\infty}
=(a(x_1)\ldots a(x_N),h)=P^h(X|).
$$
The linear independence of the polynomials $P^h(X|)$ for the basic elements of the algebra $\cA$ was proven at the end of the Subsection~\ref{th1-proof}.

Now we may apply the deformation argument. Since the map $\Phi_a$ is a bijection at one point and it is defined in terms of rational functions of $\e^{\i\pi a}$, it must be a bijection for nearly all values of the parameter~$a$. This finishes the proof.

On the physical level of strictness Theorem~\ref{theorem-bijection} has a

\begin{corollary}\label{phys-bijection}
For generic values of $a$ the expression (\ref{fhdef}) provides a one-to-one correspondence between the elements $g\in\cA\otimes\bar\cA$ and the descendant operators over the exponential field\/~$V_a(x)$. This also provides a one-to-one correspondence between the elements of the subspace $\cA_n\otimes1$ and the level $n$ chiral descendants and that between the elements of the subspace $1\otimes\bar\cA_n$ and the level $n$ antichiral descendants.
\end{corollary}

\subsection{Odd generators and integrals of motion}

The sine/h-Gordon model possesses a set of commuting integrals of motion $I_{2n-1}$ of spin $s=2n-1$ for any integer~$n$:
$$
I_{2n-1}=\int{dz\over2\pi}\,T_{2n}(x)
+\int{d\bz\over2\pi}\,\bT_{2n-2}(x).
$$
Both integrals must be taken along the same space-like contour in the $x$ plane. Some of the first currents are
$$
T_2=T^+,
\qquad
T_4=\lcolon(T^+)^2\rcolon,
\qquad
T_0=\bar T_0=-\Theta^+,
\qquad
\bT_{-2}=\bT^+,
\qquad
\bT_{-4}=\lcolon(\bT^+)^2\rcolon,
$$
where $T^+(x)$, $\bT^+(x)$, $\Theta^+(x)$ are proportional to components of the twisted energy-momentum tensor (see Eqs.~(\ref{Tcomp-def}) and (\ref{Tpmcomp-def}) in the Appendix).

Let $\cO(x)$ be any local operator. Then
\eq{
[I_{2n-1},\cO(x)]=\cI_{2n-1}\cO(x)
\equiv-\oint{dz'\over2\pi}\,T_{2n}(x')\cO(x)-\oint{d\bar z'\over2\pi}\,\bT_{2n-2}(x')\cO(x).
\label{IMcommutator}
}
The integrations here are taken over very small circles around the point~$x$ in the Euclidean plane. Hence, just the leading terms in the operator product expansions contribute to the integrals, which can be thus calculated within the conformal field theory. In particular,
$$
\Aligned{
\cI_1\cO(x)
&=-\i\cL^+_{-1}\cO(x),
&\qquad
\cI_3\cO(x)
&=-2\i\sum_{n\ge-1}\cL^+_{-n-3}\cL^+_n\cO(x),
\\
\cI_{-1}\cO(x)
&=\i\bar\cL^+_{-1}\cO(x),
&\qquad
\cI_{-3}\cO(x)
&=2\i\sum_{n\ge-1}\bar\cL^+_{-n-3}\bar\cL^+_n\cO(x).
}
$$
The operators $\cL^+_n$ acting on the space of local operators are defined as $|\cL^+_n\cO\rangle_\rad=L^+_n|\cO\rangle_\rad$ with $L^+_n$ being the standard generators of the Virasoro algebra associated with the current~$T^+(x)$ (and similar for~$\bar\cL^+_n$).

The local integrals of motion are known to be diagonalized by the many-particle states in the form
\eq{
I_{2n-1}|\theta_1,\ldots,\theta_N\rangle
=J_{|2n-1|}\sum^N_{i=1}\e^{(2n-1)\theta_i}|\theta_1,\ldots,\theta_N\rangle
\label{IMeigenvalues}
}
with some constants $J_{2n-1}$. In particular, the first integrals of motion are just the components of the momentum:
\eq{
I_1=P_z,
\qquad
I_{-1}=-P_\bz,
\qquad
J_1=-{m\over2}.
\label{Ipm1}
}
Hence, the form factors of the operator (\ref{IMcommutator}) are given by
\eq{
\langle\vac|\cI_{2n-1}\cO|\theta_1,\ldots,\theta_N\rangle
=-J_{|2n-1|}\sum^N_{i=1}\e^{(2n-1)\theta_i}\cdot
\langle\vac|\cO|\theta_1,\ldots,\theta_N\rangle.
\label{IMcOff}
}
From~(\ref{Phdef}) we get
\eq{
\Aligned{
\langle\vac|V^{c_{1-2n}g}_a|\theta_1,\ldots,\theta_N\rangle
&=\sum^N_{i=1}\e^{(2n-1)\theta_i}\cdot
\langle\vac|V^g_a|\theta_1,\ldots,\theta_N\rangle,
\\
\langle\vac|V^{\bar c_{1-2n}g}_a|\theta_1,\ldots,\theta_N\rangle
&=\sum^N_{i=1}\e^{-(2n-1)\theta_i}\cdot
\langle\vac|V^g_a|\theta_1,\ldots,\theta_N\rangle.
}\label{coddaction}
}
Comparing it with Eq.~(\ref{IMcOff}), we obtain a correspondence
\eq{
c_{1-2n}\leftrightarrow-{1\over J_{2n-1}}\cI_{2n-1},
\qquad
\bar c_{1-2n}\leftrightarrow-{1\over J_{2n-1}}\cI_{1-2n}.
\label{cIMcorresp}
}

\section{Reflection relations: basics and examples}

\subsection{General setup}

There is a conjecture based on the Liouville theory that the operators $V_a=V_{(\alpha)}$ and $V_{-a}=V_{(2\alpha_0-\alpha)}$ coincide in the breather sector up to an $\alpha$-dependent reflection factor~\cite{Zamolodchikov:1995aa}:
\eq{
\Gathered{
\langle{\rm vac}|V_a(x)|\theta_1,\ldots,\theta_N\rangle
=R_a\langle{\rm vac}|V_{-a}(x)|\theta_1,\ldots,\theta_N\rangle,
\\
R_a=m^{8\alpha_0^2a}\vR_a
=\left(m\left(p+1\over p\right)^{p+1}
{\Gamma\left(p+1\over2\right)\Gamma\left(2-p\over2\right)
\over4\sqrt\pi}
\right)^{{8\alpha_0^2a}}\>
{\Gamma\bigl(1-2{a\over p}\bigr)
\Gamma\bigl(1+2{a\over p+1}\bigr)
\over
\Gamma\bigl(1+2{a\over p}\bigr)
\Gamma\bigl(1-2{a\over p+1}\bigr)}.
}\label{Ra}
}
Since
$$
G_a=R_aG_{-a},
$$
it means that
\eq{
f_a(\theta_1,\ldots,\theta_N)=f_{-a}(\theta_1,\ldots,\theta_N).
\label{exp-reflection}
}
Later, in Subsection~\ref{exp-recurrent} we give a detailed proof of this relation.

It is natural to suppose that this correspondence extends to the whole Fock modules for generic values of~$a$. It means that for any descendant of the exponential field $V_a$ there is a unique descendant of the exponential field~$V_{-a}$ such that the multibreather matrix elements of these two operators coincide. For particular values of $a$ this correspondence may look broken, but it must be recoverable by an appropriate limiting procedure. Sometimes, it demands extending the Fock modules by the action of the operator~$\bfQ$. With our Conjecture~\ref{phys-bijection} this correspondence means that the reflection relations map each element $g\in\cA\otimes\bar\cA$ on an element $g'$ so that the form factors $f^g_a=f^{g'}_{-a}$ and establish an $a$-dependent family of bijections in the space $\cA\otimes\bar\cA$. Note that, due to the cluster property~(\ref{cluster}), this map, if exists, possesses a factorized form on the tensor product $\cA\otimes\bar\cA$.

\begin{theorem}[Reflection property]\label{reflectiontheorem}
For generic values of the parameter $a$ there exists a linear automorphism $r_a:\cA\to\cA$, such that
\eq{
f^{h\bar h'}_a(\theta_1,\ldots,\theta_N)
=f^{r_a(h)\overline{r_{-a}(h')}}_{-a}
(\theta_1,\ldots,\theta_N).
\label{reflection}
}
The automorphism $r_a$ admits a restriction to an automorphism of each of the subspaces~$\cA_n$.
\end{theorem}

We defer the proof of the theorem to the Section~\ref{reflectionproof}. In this section we describe some properties of this bijection and obtain an example of reflection relations by `handicraft' methods.

Besides, there is another relation for the form factors with different values of~$a$, which is an evident consequence of the definition (\ref{cTdef}),~(\ref{fhdef}):
\eq{
f^g_{a+1}(\theta_1,\ldots,\theta_N)
=(-)^Nf^g_a(\theta_1,\ldots,\theta_N)
\label{fperiodicity}
}
for an arbitrary element $h\in\cA\otimes\bar\cA$. We shall make use of this relation below.

Consider any basis in each $\cA_n$: $h^{[n,i]}$, $(i=1,\ldots,\dim\cP_n)$. Let $\bar h^{[n,i]}$ be the respective basis in~$\bar\cA$. There is a set of form factors
$$
f^{[n,i][\bar n,j]}_a(\theta_1,\ldots,\theta_N)
\equiv f^{h^{[n,i]}\bar h^{[n,i]}}_a(\theta_1,\ldots,\theta_N).
$$
The property (\ref{fperiodicity}) is given by
\eq{
f^{[n,i][\bar n,i']}_{a+1}(\theta_1,\ldots,\theta_N)
=(-)^Nf^{[n,i][\bar n,i']}_a(\theta_1,\ldots,\theta_N).
\label{fperiodicity-b}
}

The reflection property can be formulated as follows. There exists an $a$ dependent but $N$ independent matrix $U^{[n]}_a=(U^{[n,ij]}_a)_{i,j=1}^{\dim\cP_n}$ such that
\eq{
f^{[n,i][\bar n,i]}_a(\theta_1,\ldots,\theta_N)
=\sum_{j,j'}U^{[n,ij]}_a U^{[\bar n,i'j']}_{-a}
f^{[n,j][\bar n,j']}_{-a}(\theta_1,\ldots,\theta_N).
\label{reflection-b}
}

It is only necessary to prove the conjecture for the products of the elements $c_{-2m}$, because the polynomials corresponding to the elements $c_{-2m+1}$ factor out from the expressions for form factors.

Let us try to check this property on the level~$(2,0)$. The space $\cP_2$ is two-dimensional. There are two basic elements of the algebra on the level 2, $c_{-1}^2$ and~$c_{-2}$. The corresponding basic polynomials are (I omit the superscript corresponding to the antichiral sector)
$$
\Aligned{
P^{[2,1]}_{N,k}(X|Y)
&=\left(\sum x_i+\sum y_i\right)^2,
\\
P^{[2,2]}_{N,k}(X|Y)
&=\sum x_i^2-\sum y_i^2.
}
$$
The corresponding two-particle $J$ functions are given by
\subeq{\Align{
J^{[2,1]}_{2,a}(x_1,x_2)
&=4\cos^2\pi a\cdot(x_1+x_2)^2,
\label{f[2,1]}
\\*
J^{[2,2]}_{2,a}(x_1,x_2)
&=2\i\sin{2\pi a}\>(x_1+x_2)^2+4\i\left(\sin\pi p-\sin2\pi a\right)x_1x_2.
\label{f[2,2]}
}\label{f[2]}}
The matrix $U^{[2]}_a$ looks in this bases unnaturally complicated. But the first term in (\ref{f[2,2]}) coincides with (\ref{f[2,1]}) up to a constant factor. If we subtracted it, the $U$ matrix would be diagonal. Then define
\subeq{\Align{
f^{(1,1)}_a(\theta_1,\ldots,\theta_N)
&=f^{[2,1]}_a(\theta_1,\ldots,\theta_N),
\label{f(1,1)}
\\*
f^{(2)}_a(\theta_1,\ldots,\theta_N)
&={1\over\sin\pi p-\sin2\pi a}
\left(
f^{[2,2]}_a(\theta_1,\ldots,\theta_N)
-\i f^{[2,1]}_a(\theta_1,\ldots,\theta_N)\tg\pi a
\right).
\label{f(2)}
}\label{f()}}
For $N=2$ we have
$$
\Aligned{
J^{(1,1)}_{2,a}(x_1,x_2)
&=4\cos^2\pi a\>(x_1+x_2)^2,
\\
J^{(2)}_{2.a}(x_1,x_2)
&=4\i\,x_1x_2.
}
$$
The corresponding form factors possess the property
$$
\Gathered{
f^{(1,1)}_a(\theta_1,\theta_2)
=f^{(1,1)}_{-1-a}(\theta_1,\theta_2)
=f^{(1,1)}_{-a}(\theta_1,\theta_2),
\\
f^{(2)}_a(\theta_1,\theta_2)
=f^{(2)}_{-1-a}(\theta_1,\theta_2)
=f^{(2)}_{-a}(\theta_1,\theta_2).
}
$$
We may conjecture that for arbitrary~$N$
\subeq{\Gather{
f^{(1,1)}_a(\theta_1,\ldots,\theta_N)
=(-)^Nf^{(1,1)}_{-1-a}(\theta_1,\ldots,\theta_N)
=f^{(1,1)}_{-a}(\theta_1,\ldots,\theta_N),
\label{f(1,1)reflection}
\\
f^{(2)}_a(\theta_1,\ldots,\theta_N)
=(-)^Nf^{(2)}_{-1-a}(\theta_1,\ldots,\theta_N)
=f^{(2)}_{-a}(\theta_1,\ldots,\theta_N).
\label{f(2)reflection}
}\label{f()reflection}}
The equation (\ref{f(1,1)reflection}) is a direct consequence of~(\ref{exp-reflection}). Eq.~(\ref{f(2)reflection}) will be proven in Subsection~\ref{(2,0)proof}.

In the algebraic language we defined two elements
\Align{
h^{(1,1)}_a
&=c_{-1}^2,
\label{h(1,1)}
\\*
h^{(2)}_a
&={c_{-2}-\i c_{-1}^2\tg\pi a\over\sin\pi p-\sin2\pi a}.
\label{h(2)}
}
The reflection map on the level $(2,0)$ is thus given by
\eq{
r_a(h^{(1,1)}_a)=h^{(1,1)}_{-a},
\qquad
r_a(h^{(2)}_a)=h^{(2)}_{-a}.
\label{reflection(2,0)}
}

The denominator of~(\ref{h(2)}) possesses two zeros (up to periodicity) at the points $a=p/2$ and $a=-(1+p)/2$, which corresponds to $\alpha=\alpha_+/2$ and $\alpha=\alpha_-/2$. Note that these points correspond to two modules of one of the Virasoro algebras (corresponding to the current $T^-$ defined in~(\ref{Tpm-def}) below) degenerate at the level two. From (\ref{reflection(2,0)}) we have
$$
f^{c_{-2}-\i c_{-1}^2\tg\pi a}_a(\theta_1,\ldots,\theta_N)
={\sin\pi p-\sin2\pi a\over\sin\pi p+\sin2\pi a}
f^{c_{-2}+\i c_{-1}^2\tg\pi a}_a(\theta_1,\ldots,\theta_N).
$$
But the form factor in the right hand side is surely finite everywhere in $a$ including the points $a=p/2,-(1+p)/2$. Hence, for these particular values of $a$ the form factor of a finite element vanishes:
$$
f^{c_{-2}-\i c_{-1}^2\tg\pi a}_a(\theta_1,\ldots,\theta_N)=0
\quad\text{for $a=p/2$ or $a=-(1+p)/2$.}
$$
It breaks the bijection property at these points. The finite form factors correspond to the element $h^{(2)}_a$ which is undefined there. The corresponding functions $P^{h^{(2)}_a}_a$ are also undefined. Nevertheless, the form factors are well-defined as limits. We expect that for generic $p$ this break of the bijection property takes place at the values
$$
\alpha=\mp\alpha_{kl},
\qquad
\alpha_{kl}={1-k\over2}\alpha_++{1-l\over2}\alpha_-,
\qquad
k,l=1,2,\ldots,
$$
which correspond to degeneration points of the Fock module as a representation of the Virasoro algebras generated by $T^-(x)$ and $\bar T^+(x)$. The bijection breaks starting from the level~$kl$ in the chiral sector for the minus sign and in the antichiral sector for the plus sign.

The families of elements $h^{(1,1)}_a$ and $h^{(2)}_a$ are `self-dual' in the sense of~(\ref{reflection(2,0)}). More generally, if the matrices $U^{[n]}_a$ admit analytic factorization,
$$
U^{[n]}_a=\left(W^{[n]}_{-a}\right)^{-1}W^{[n]}_a,
\qquad
W^{[n]}_{a+1}=W^{[n]}_a,
$$
one can define `self-dual' bases of in the spaces~$\cA_n$,
$$
h^{(n,i)}_a=\sum_{i'}W^{[n,ii']}h^{[n,i']},
\qquad
r_a(h^{(n,i)}_a)=h^{(n,i)}_{-a}.
$$
The corresponding `self-dual' form factors
$$
f^{(n,i;\bar n,j)}_a(\theta_1,\ldots,\theta_N)
=\sum_{i'j'}W^{[n,ii']}_aW^{[\bar n,jj']}_{-a}f^{[n,i'][\bar n,j']}_a(\theta_1,\ldots,\theta_N)
$$
satisfy the equations
$$
f^{(n,i;\bar n,j)}_a(\theta_1,\ldots,\theta_N)
=(-)^Nf^{(n,i;\bar n,j)}_{-1-a}(\theta_1,\ldots,\theta_N)
=f^{(n,i;\bar n,j)}_{-a}(\theta_1,\ldots,\theta_N).
$$

\subsection{Analytic properties of the $J^g_{N,a}$ functions}

In this section we set $I=(1,\ldots,N')$, $X=(x_1,\ldots,x_{N'})$, where the value of $N'\le N$ will be always clear from the context. Besides, we use the notation $\hat I_i=I\setminus\{i\}$, $\hat X_i=X\setminus\{x_i\}$.

First, let us prove that the function $J^g_a(x_1,\ldots,x_N)$ is regular on the hyperplanes $x_i=x_j$. The proof is straightforward. The contribution to the corresponding residue comes from the terms for which either $i\in I_-$, $j\in I_+$ or $i\in I_+$, $j\in I_-$. Let $i=N-1$, $j=N$. Then
$$
\Aligned{
\Res_{x'=x}J^g_{N,a}(X,x,x')
&=\sum_{I_++I_-=I}
\e^{\i\pi a(\#I_--\#I_+)}P^g(X_-,x|X_+,x)
\\
&\quad\times
\prod_{i\in I_-\atop j\in I_+}f\left(x_i\over x_j\right)
\prod_{i\in I_-}f\left(x_i\over x_\vpj\right)
\prod_{j\in I_+}f\left(x\over x_j\right)
\\
&\quad\times
x(\Res_{z=1}f(z)+\Res_{z=1}f(1/z))=0.
}
$$

Second, let us find the residues in $x_i$ at the points $x_i=-x_j$. Using the property $f(-x)=f(x^{-1})$, we obtain
\eq{
\Res_{x'=-x}J^g_{N,a}(X,x,x')
=\i x\sin\pi p\cdot
\left(\prod^{N-2}_{i=1}f\left(x_i\over x_\vpj\right)
-\prod^{N-2}_{i=1}f\left(x\over x_i\right)\right)
J^g_{N,a}(X).
\label{Jhred}
}
Introduce the functions
\Align{
R^g_{N,a,i}(X)
&={1\over x_i}\Res_{x=-x_i}J^g_{N,a}(X,x)
\notag
\\
&=-\i\sin\pi p\cdot
\left(\prod_{j\in\hat I_i}f\left(x_i\over x_j\right)
-\prod_{j\in\hat I_i}f\left(x_j\over x_i\right)\right)
J^g_{N-2,a}(\hat X_i).
\label{Ridef}
}
We shall also use the notation
\eq{
D^g_{N,a}(X)=\sum_{i=1}^{N-1}R^g_{N,a,i}(X).
\label{sumR}
}

Consider the function $J^g_{N,a}(X,x)$ as an analytic function of the variable $x$ depending on the parameter $X=(x_1,\ldots,x_{N-1})$. We may separate the contribution of the poles at the points $x=-x_i$:
\eq{
J^g_{N,a}(X,x)
=\sum_{i=1}^{N-1}{x_iR^g_{N,a,i}(X)\over x+x_i}
+J^{(\infty)g}_{N,a}(X,x)
\label{Jinfform}
}
The function $J^{(\infty)g}_{N,a}(X,x)$ is regular everywhere except the points $x=0,\infty$. Since the sum over the residues here is of the order $O(x^{-1})$ as $x\to\infty$, the asymptotic behavior of $J^g$ as a function of $x$ is governed by $J^{(\infty)g}$:
$$
J^g_{N,a}(X,x)-J^{(\infty)g}_{N,a}(X,x)=O(x^{-1})
\quad\text{as $x\to\infty$}.
$$
With the notation (\ref{sumR}) we have
\eq{
\sum_{i=1}^{N-1}{x_iR^g_{N,a,i}(X)\over x+x_i}
=D^g_{N,a}(X)-\sum_{i=1}^{N-1}{x_i^{-1}R^g_{N,a,i}(X)\over x^{-1}+x_i^{-1}}
\label{Ridentity}
}
and
\Gather{
J^g_{N,a}(X,x)
=-\sum_{i=1}^{N-1}{x_i^{-1}R^g_{N,a,i}(X)\over x^{-1}+x_i^{-1}}
+J^{(0)g}_{N,a}(X,x),
\label{J0form}
\\
J^{(0)g}_{N,a}(X,x)=J^{(\infty)g}_{N,a}(X,x)+D^g_{N,a}(X).
\label{J0infrel}
}
It is evident that the behavior of the function $J^g_{N,a}(X,x)$ in the vicinity of the singularity $x=0$ is governed by $J^{(0)g}_{N,a}(X,x)$:
$$
J^g_{N,a}(X,x)-J^{(0)g}_{N,a}(X,x)=O(x)
\quad\text{as $x\to0$}.
$$
It means the functions $J^{(\infty)g}_{N,a}(X,x)$ and $J^{(0)g}_{N,a}(X,x)$ as functions of $x$ have the only singularities at the points $x=0$, $x=\infty$ and their singular parts coincide with those of the functions $J^g_{N,a}(X,x)$ in the vicinity of both singularities. It is easy to see that the singularities are in fact poles. If $g=h\bar h'$ is a homogeneous element of $\cA\otimes\bar\cA$ of the order $(n,\bar n)$, then
\subeq{\Align{
J^{(\infty)h\bar h'}_{N,a}(X,x)
&=O(x^n)\quad\text{as $x\to\infty$},
\label{Jinfbound}
\\
J^{(0)h\bar h'}_{N,a}(X,x)
&=O(x^{-\bar n})\quad\text{as $x\to0$}.
\label{J0bound}
}\label{Jbound}}
It means that $J^{(\infty)}$, $J^{(0)}$ are Laurent polynomials of~$x$:
\eq{
J^{(\infty/0)h\bar h'}_{N,a}(X,x)=\sum_{s=-\bar n}^n C^{(\infty/0)h\bar h'}_{N,a,s}(X)x^s,
\qquad
C^{(0)h\bar h'}_{N,a,s}(X)-C^{(\infty)h\bar h'}_{N,a,s}(X)=D^{h\bar h'}_{N,a}(X)\delta_{s,0}.
\label{Jinf0expand}
}
Note that the lowest and coefficient, $C^{h\bar h'}_{N,a,-\bar n}$, and the highest one, $C^{h\bar h'}_{N,a,n}$, are fixed by the cluster property~(\ref{cluster}):
\eq{
C^{(\infty)h\bar h'}_{N,a,n}(X)=J^{\bar h'}_{N-1,a}(X)\cdot\left.x^{-n}J^h_{1,a}(x)\right|_{x\to\infty},
\qquad
C^{(0)h\bar h'}_{N,a,-\bar n}=J^h_{N-1,a}(X)\cdot\left.x^{\bar n}J^{\bar h'}_{1,a}(x)\right|_{x\to0}.
\label{Chl}
}

Since $D^g_{N,a}(X)$ is expressed in terms of the functions $J^g_{N-2,a}$ we would express all form factors recursively in $N$ if we could express $J^{(\infty)g}_{N,a}(X,x)$ in terms of $J^g_{N-1,a}(X)$. Up to now this problem is solved in very few cases. They will be described in the rest of this section.

\subsection{Recurrent relations and reflection property: exponential fields}
\label{exp-recurrent}

Consider the simplest case of exponential fields, $g=1$ (we shall always omit the superscript $g$ if it is equal to one). We have
$$
J^{(\infty)}_{N,a}(X,x)=\const_x,
\qquad
J^{(0)}_{N,a}(X,x)=\const_x.
$$
To fix the constants we have to calculate $J_{N,a}(X,0)$ and $J_{N,a}(X,\infty)$. Since $f(0)=f(\infty)=1$, we obtain
$$
J_{N,a}(X,0)=J_{N,a}(X,\infty)=2\cos\pi a\cdot J_{N-1,a}(X)
$$
and, hence,
\eq{
J^{(\infty)}_{N,a}(X,x)=J^{(0)}_{N,a}(X,x)=2\cos\pi a\cdot J_{N-1,a}(X).
\label{J0inf}
}
The fact that $D_{N,a}(X)=0$ provides a nontrivial identity
\eq{
\sum^{N-1}_{i=1}R_{N,a,i}(X)=0
\label{sumRprim}
}
It is characteristic for the exponential fields.

We arrive to the

\begin{theorem}

The recurrent relations
\Align{
J_{N,a}(X,x)
&=2\cos\pi a\cdot J_{N-1,a}(X)
+\sum^{N-1}_{i=1}{x_iR_{N,a,i}(X)\over x+x_i},
\label{gNNrec}
}
where
\eq{
R_{N,a,i}(X)
=-\i\sin\pi p\cdot
\left(\prod_{j\in\hat I_i}f\left(x_i\over x_j\right)
-\prod_{j\in\hat I_i}f\left(x_j\over x_i\right)\right)
J_{N-2,a}(\hat X_i),
\label{R1def}
}
with the initial conditions
\eq{
J_{0,a}=1,
\qquad
J_{1,a}(x)=2\cos\pi a
\label{grecinit}
}
define uniquely a set of homogeneous symmetric functions $J_{N,a}$ of $N$ variables of partial power~$0$.

\end{theorem}

From this relation we readily get
$$
J_{N,a}(x_1,\ldots,x_N)=J_{N,-a}(x_1,\ldots,x_N).
$$
This proves~(\ref{exp-reflection}).

Note that the described recurrent relation is an explicit form of the relation found implicitly (in the form of a uniqueness theorem) by Koubek and Mussardo~\cite{Koubek:1993ke}.

\subsection{Recurrent relation and reflection property: level $\>(2,0)\!$ descendants}
\label{(2,0)proof}

Let us prove the relations (\ref{f(2)reflection}) or, equivalently, the second equation of~(\ref{reflection(2,0)}). We want to turn the relation (\ref{Jinfform}) or (\ref{J0form}) into a recursive relation. We know all the residues at the poles $x=-x_i$ ($i=1,\ldots,N-1$) due to (\ref{Ridef}) and the leading coefficients in the asymptotics $C^{(\infty)h}_{N,a,2}$, $C^{(0)h}_{N,a,0}$ ($h\in\cA_2$) due to~(\ref{Chl}). The only thing we need to find is the coefficient $C^{(\infty)h}_{N,a,1}=C^{(0)h}_{N,a,1}$ of the subleading term. Hence, let us try to expand the expression for $J^h_{N,a}(X,x)$ in $x$ up to the terms of the order $x^1$ as $x\to\infty$.

Consider first the element~$c_{-2}$ and separate in the expression for $J^{c_{-2}}_{N,a}(x_1,\ldots,x_{N-1},x)$ the terms containing $x^2$ in the polynomials $P^{c_{-2}}(X_-|X_+)$:
$$
J^{c_{-2}}_{N,a}(X,x)
=x^2 K^{(-)}_{N,a}(X|x)+K^{(+)c_{-2}}_{N,a}(X|x),
$$
where
$$
\Aligned{
K^{(\pm)g}_{N,a}(X|x)
&=\sum_{I_++I_-=\hat I_N}\e^{\i\pi a(\#I_--\#I_+)}
P^g(X_-|X_+)\prod_{i\in I_-\atop j\in I_+}f\left(x_i\over x_j\right)
\\
&\quad\times
\left(\e^{\i\pi a}\prod_{j\in I_+}f\left(x\over x_j\right)
\pm\e^{-\i\pi a}\prod_{i\in I_-}f\left(x_i\over x_\vpj\right)
\right).
}
$$
Note that $K^{(+)}_{N,a}(X|x)=J_{N,a}(X,x)$.

It is evident that the functions $K^{(\pm)g}_{N,a}$ are of the order~$O(x^0)$ as $x\to\infty$. It means that we may completely ignore the function $K^{(+)c_{-2}}$ while considering the asymptotics of $J^{c_{-2}}_{N,a}(X,x)$ with the accuracy~$O(x^0)$.

Since $f(x)=1+x^{-1}\cdot2\i\sin\pi p+O(x^{-2})$, we have
$$
K^{(-)}_{N,a}(X|x)=J_{N-1,a}(X)\cdot2\i\sin\pi a
+{2\i\sin\pi p\over x}L_{N-1,a}(X)+O\left(1\over x^2\right).
$$
Here
$$
\Aligned{
L_{N,a}(X)
&=\sum_{I_++I_-=I}\e^{\i\pi a(\#I_--\#I_+)}
\left(
\e^{\i\pi a}\sum_{j\in I_+}x_j+\e^{-\i\pi a}\sum_{i\in I_-}x_i
\right)
\prod_{i\in I_-\atop j\in I_+}f\left(x_i\over x_j\right)
\\
&=J^{c_{-1}}_{N,a}(X)\cos\pi a-\i L'_{N,a}(X)\sin\pi a,
\\
L'_{N,a}(X)
&=\sum_{I_++I_-=I}\e^{\i\pi a(\#I_--\#I_+)}
\left(
\sum_{i\in I_-}x_i-\sum_{j\in I_+}x_j
\right)
\prod_{i\in I_-\atop j\in I_+}f\left(x_i\over x_j\right)
}
$$
Finally,
\Align{
J^{c_{-2}}_{N,a}(X,x)
&=x^2J_{N-1,a}(X)\cdot2\i\sin\pi a
\notag
\\*
&\quad+x(J^{c_{-1}}_{N-1,a}(X)\cos\pi a-\i L'_{N-1,a}(X)\sin\pi a)
\cdot2\i\sin\pi p+O(x^0)
\label{Jc-2asymp}
}
as $x\to\infty$. Similarly, we have
\Align{
J^{c_{-1}^2}_{N,a}(X,x)
&=x^2K^{(+)}_{N,a}(X|x)+2xK^{(+)c_{-1}}_{N,a}(X|x)
+K^{(+)c_{-1}^2}_{N,a}(X|x)
\notag
\\*
&=x^2J_{N-1,a}(X)\cdot2\cos\pi a
\notag
\\*
&\quad
+2x((J^{c_{-1}}_{N-1,a}(X)(2\cos\pi a-\sin\pi a\sin\pi p)
-\i L'_{N-1,a}(X)\cos\pi a\sin\pi p)+O(x^0).
\label{Jc-1^2asymp}
}
It is surely possible to study the asymptotics of the functions $L'_{N,a}$, but we do not need it. Choose such a linear combination of (\ref{Jc-2asymp}) and (\ref{Jc-1^2asymp}) that the terms containing $L'_{N-1,a}(X)$ cancel each other in it. It is
$$
J^{c_{-2}}_{N,a}(X,x)-\i J^{c_{-1}^2}_{N,a}(X,x)\tg\pi a
=2\i x{\sin\pi p-\sin2\pi a\over\cos\pi a}J^{c_{-1}}_{N-1,a}(X)
+O(x^0).
$$
Dividing it by $\sin\pi p-\sin2\pi a$ we get
$$
J^{h^{(2)}_a}_{N,a}(X,x)
=x{2\i\over\cos\pi a}J^{c_{-1}}_{N-1,a}(X)+O(x^0)
\quad\text{as $x\to\infty$}.
$$
The value at $x=0$ is known from~(\ref{Chl}):
$$
J^{(0)h^{(2)}_a}_{N,a}(X,0)
=C^{(0)h^{(2)}_a}_{N,a,0}=2\cos\pi a\cdot J^{h^{(2)}_a}_{N-1,a}(X).
$$
Therefore
\eq{
J^{(0)h^{(2)}_a}_{N,a}(X,x)
=x{2\i\over\cos\pi a}J^{c_{-1}}_{N-1,a}(X)+2\cos\pi a\cdot J^{h^{(2)}_a}_{N-1,a}(X).
\label{J(0)h2}
}
Unlike the situation in the case of the primary fields, the function $D^{h^{(2)}_a}_{N,a}(X)$ is not fixed by the asymptotics and can be calculated directly from the definition~(\ref{sumR}) only.

As a result, the expression (\ref{J0form}) for $h=h^{(2)}_a$ together with (\ref{Ridef}) and (\ref{J(0)h2}) becomes a recurrent relation for the functions~$J^{h^{(2)}_a}_{N,a}(X)$. This relation with the initial conditions $J^{h^{(2)}_a}_{0,a}=J^{h^{(2)}_a}_{1,a}(x)=0$ defines $J^{h^{(2)}_a}_{N,a}(X)$ uniquely. Again, the parameter $a$ always enters the recurrent relations and initial conditions in the combination $\cos\pi a$, which proves~(\ref{reflection(2,0)}). Note that the first term of (\ref{J(0)h2}) is finite at $a=\pm1/2$ since the zeros of the function $J^{c_1}_{N-1,a}$ at these points cancel the poles arising from $\cos\pi a$ in the denominator. Hence, the resulting $J$ functions are well-defined at these points.

In principle, it is possible to construct the recurrent relations at each level. Nevertheless, even in the case of level~$(2,2)$ descendants they become enormous. They seem to give no chance to prove the reflection property in general. For this reason we develop another approach to prove the existence of the reflection relations in general, based on the expansions of the form factors of exponential operators. Technically, it uses a novel two-boson representation described below. Some applications of the recursion relations are collected in the Appendix.

\section{A free field representation for the functions $J_{N,a}^g(X)$}
\label{ddffrep}

According to Eq.~(\ref{fJ}) each form factor is proportional to the functions $J_{N,a}^g(X)$ up to a factor uniform for all form factors with given number of particles. On the other hand, the expression (\ref{Jadef}) looks like a matrix element of a combination of vertex operators. In this section we define these vertex operators in terms of free fields.

Consider the Heisenberg algebra generated by the elements $d^\pm_n$ $(n\in\Z)$ with the commutation relations:
\eq{
[d^\pm_m,\hat a]=0,
\qquad
[d^\pm_m,d^\pm_n]=0,
\qquad
[d^\pm_m,d^\mp_n]=mA^\pm_m\delta_{m+n,0}.
\label{ddcommut}
}
Here
\eq{
\Aligned{
A^+_{2k}
&=A^-_{2k}=(\omega^k-\omega^{-k})^2=-4\sin^2\pi kp,
\\
A^+_{2k-1}
&=-A^-_{2k-1}=\omega^{2k-1}-\omega^{1-2k}=2\i\sin\pi(2k-1)p.
}\label{Apmdef}
}
Add a central element $\hat a$ and define the vacuums
\eq{
d^\pm_n|1\rangle_a=0,
\quad
\hat a|1\rangle_a=a|1\rangle_a,
\quad
{}_a\langle1|d^\pm_{-n}=0,
\quad
{}_a\langle1|\hat a={}_a\langle1|a,
\quad
{}_a\langle1|1\rangle_a=1\quad(n>0).
\label{dvacuum}
}
Introduce the vertex operators
\eq{
\lambda_\pm(z)=\lcolon\exp\left(\sum_{n\ne0}{d^\pm_n\over n}z^{-n}\right)\rcolon.
\label{lambdapmdef}
}
These vertex operators satisfy the relations
\subeq{\Align{
\lambda_\pm(z')\lambda_\pm(z)
&=\lcolon\lambda_\pm(z')\lambda_\pm(z)\rcolon,
\label{lambdappmmprod}
\\*
\lambda_+(z')\lambda_-(z)
&=\lambda_-(z)\lambda_+(z')
=f\left(z\over z'\right)\lcolon\lambda_+(z')\lambda_-(z)\rcolon.
\label{lambdapmmpprod}
}\label{lambdaprods}}
The operators $\lambda_+(z')$ and $\lambda_-(z)$ commute everywhere except the points $z'=\pm z$. Define a combination
\eq{
t(z)=\e^{\i\pi\hat a}\lambda_-(z)+\e^{-\i\pi\hat a}\lambda_+(z).
\label{tzdef}
}
It looks much the same as~(\ref{Tdef}). Then
\eq{
J_{N,a}(X)=\langle t(x_1)\ldots t(x_N)\rangle_a.
\label{Jtrel}
}
The whole construction looks similar to the standard free field representation~\cite{Lukyanov:1997bp}, but there are several differences. The set of the oscillators here is countable rather than continuous and the form factor is proportional to a vacuum expectation rather than to a trace. Besides, it strips off the annoying factors $R(\theta_i-\theta_j)$. These are advantages. The price we pay for these simplifications is that the residue of the kinematic pole of the vertex operators is already not a $c$-number but a new vertex operator (see (\ref{s(z)def}) below). We shall see that this new vertex will play an important role in the proof of the reflection relations.

Now introduce two homomorphisms of the algebra~$\cA$ into the Heisenberg algebra:
\eq{
\pi_R(c_{-n})={d^-_n-d^+_n\over A^+_n},
\qquad
\pi_L(c_{-n})={d^-_{-n}-d^+_{-n}\over A^+_n}
\qquad
(n>0).
\label{piLRdef}
}
It is easy to check that
\eq{
\Aligned{{}
[\pi_R(c_{-n}),\lambda_\pm(z)]
&=(\mp)^{n+1}z^n\lambda_\pm(z),\qquad&
\pi_R(c_{-n})|1\rangle_a
&=0,
\\
[\pi_L(c_{-n}),\lambda_\pm(z)]
&=-(\pm)^{n+1}z^{-n}\lambda_\pm(z),\qquad&
{}_a\langle1|\pi_L(c_{-n})
&=0.
}\label{piLRlambdapmcommut}
}
Besides,
\eq{
[\pi_R(c_{-m}),\pi_L(c_{-n})]=\delta_{m,n}\times
\Cases{0,&m\in2\Z+1,\\
-2m(A^+_m)^{-1},&m\in2\Z.}
\label{piLRcommut}
}
The maps $\pi_R$ and $\pi_L$ may be considered as a right and a left representation of the algebra~$\cA$. Let
\eq{
{}_a\langle h|={}_a\langle1|\pi_R(h),
\qquad
|\bar h\rangle_a=\pi_L(h)|1\rangle_a.
\label{hstatesdef}
}
From the equations (\ref{piLRlambdapmcommut}) we easily get
\eq{
J^h_{N,a}(X)={}_a\langle h|t(x_1)\ldots t(x_N)|1\rangle_a,
\qquad
J^{\bar h}_{N,a}(X)={}_a\langle1|t(x_1)\ldots t(x_N)|\bar h\rangle_a,
\qquad
h\in\cA.
\label{Jh1def}
}
Define also functions
\eq{
\tJ^{h\bar h'}_{N,a}(X)
={}_a\langle h|t(x_1)\ldots t(x_N)|\bar h'\rangle_a.
\label{Jtildedef}
}
For generic elements $h$ and $h'$ these functions do not coincide with the functions $J^{h\bar h'}_{N,a}(X)$, but they are related to them. Let us introduce two maps
\eq{
\pi_{LR}(h\bar h')=\pi_L(h')\pi_R(h),
\qquad
\pi_{RL}(h\bar h')=\pi_R(h)\pi_L(h').
\label{piRLLR}
}
These maps are bijections of $\cA\otimes\cA$ to the subalgebra of the Heisenberg algebra generated by the elements $d_n^--d_n^+$, $n\ne0$.
Then
\eq{
\tJ^{h\bar h'}_{N,a}(X)
=J^{\pi_{LR}^{-1}\circ\pi_{RL}(h\bar h')}_{N,a}(X).
\label{JtildeJrel}
}
More explicitly, suppose that $h=\prod^k_{\mu=1}h_\mu$, $h'=\prod^{k'}_{\nu=1}h'_\nu$, where $h_\mu$, $h'_\nu$ are any linear combinations of the generators $c_{-n}$. Then the relation is given by a Wick-type formula:
\eq{
\Gathered{
\tJ^{h\bar h'}_{N,a}(X)
=\sum^{\min(k,k')}_{l=0}
\sum^k_{\mu_1\ne\ldots\ne\mu_l}\sum^{k'}_{\nu_1<\ldots<\nu_l}
J^{h_{\smash{[\mu_1\ldots\mu_l]}}
\overline{h'_{\smash{[\nu_1\ldots\nu_l]}}}}_{N,a}(X)
\prod^l_{s=1}[\pi_R(h_{\mu_s}),\pi_L(h'_{\nu_s})],
\\
h_{[\mu_1\ldots\mu_l]}=\prod^k_{\mu\ne\mu_1,\ldots,\mu_l}h_\mu,
\qquad
h'_{[\nu_1\ldots\nu_l]}=\prod^{k'}_{\nu\ne\nu_1,\ldots,\nu_l}h'_\nu.
}\label{JtildeJexplicit}
}
The set of form factors $G_a\tf^g_a$ proportional to the function $\tJ^g_{N,a}(X)$ corresponds to a field
\eq{
\tV^g_a(x)=V^{\pi_{LR}^{-1}\circ\pi_{RL}(g)}_a(x).
\label{VtildeVrel}
}

\begin{conjecture}\label{Jtildeconjecture}
The operators $\tV^g_a(x)$ with a homogeneous element $g$ of the order $(n,\bar n)$ are descendants of the operator $V_a(x)$ of the definite level $(n,\bar n)$.
\end{conjecture}

Up to now, we are only able to substantiate this conjecture in the case of the level $(2,2)$ descendants. Due to the resonant pole on the level $(2,2)$ at $\alpha=-\beta/2$ there exists an operator $W_{(\alpha)}(x)$ at this level such that~\cite{Fateev:1998xb}
$$
\Res_{\alpha=-\beta/2}W_{(\alpha)}(x)=V_{(3\beta/2)}(x).
$$
Let
$$
\tV^{(A,\bar B)}_a(x)=\tV^{h^{(A)}_a\bar h^{(B)}_{-a}}_a(x).
$$
Consider the operator $\tV^{(2,\bar2)}_a(x)$ corresponding to the function $\tJ^g_{N,a}(X)$ with $g=h^{(2)}_a\bar h^{(2)}_{-a}$. We check by a direct calculation for $N\le4$ that indeed
$$
{\pi\sin^2\pi p\sin2\pi p\over\alpha_0}
\Res_{\alpha=-\beta/2}\left(\tJ^{h^{(2)}_a\bar h^{(2)}_{-a}}_{N,a}(X)\,d\alpha\right)
=J_{N,(3\beta/2)}(X).
$$

For general values of $a$ the descendant operator $\cL^+_{-2}\bar\cL^+_{-2}\e^{\i\alpha\varphi}$ must be, according to our conjecture, a linear combination
$$
\cL^+_{-2}\bar\cL^+_{-2}\e^{\i\alpha\varphi}
=H^{(2,\bar2)}_a\tV^{(2,\bar2)}_a+H^{(2,\overline{1,1})}_a\tV^{(2,\overline{1,1})}_a
+H^{(1,1,\bar2)}_a\tV^{(1,1,\bar2)}_a+H^{(1,1,\overline{1,1})}_a\tV^{(1,1,\overline{1,1})}_a
$$
with some coefficients $H^{(A,\bar B)}_a$. Evidently, the only contribution to the expectation comes from the first term. The coefficient $H_a\equiv H^{(2,\bar2)}_a$ should satisfy the equations~\cite{Fateev:1998xb}
$$
H_a=H_{-a},
\qquad
H_a=H_{-1-a}\left((\alpha-\alpha_{12})(\alpha-\alpha_{21})
\over(\alpha+\alpha_{12})(\alpha+\alpha_{21})\right)^2.
$$
We have to choose a solution regular at the points~$\alpha=\pm\alpha_{12},\pm\alpha_{21}$. On the other hand, the vacuum expectation value of the operator associated to $\tV^{(2,\bar2)}_a$ is equal to
$$
G_a\tilde f^{h^{(2)}_a\bar h^{(2)}_{-a}}_a()
=G_a\tilde J^{h^{(2)}_a\bar h^{(2)}_{-a}}_{0,a}
={G_a\over\sin^2\pi p\,(\sin^2\pi p-\sin^22\pi a)}.
$$
The product $H_a\tilde f^{h^{(2)}_a\bar h^{(2)}_{-a}}_a()$, properly normalized, coincides with the function ${\cal W}(-a/2)$ in~\cite{Fateev:1998xb}, which means that Conjecture~\ref{Jtildeconjecture} is consistent with the known exact expectation values found there.

\section{Proof of the reflection property}
\label{reflectionproof}

Introduce a current
\eq{
s(z)=\lcolon\lambda_-(z)\lambda_+(-z)\rcolon.
\label{s(z)def}
}
It is easy to check that
$$
s(z)t(x)=t(x)s(z)=f\left(z\over x\right)\lcolon s(z)t(x)\rcolon
$$
and
$$
\langle t(x_1)\ldots t(x_K)s(y_1)\ldots s(y_L)\rangle_a
=\prod^K_{i=1}\prod^L_{j=1}f\left(y_j\over x_i\right)
\prod^L_{j<j'}f\left(y_j\over y_{j'}\right)f\left(y_{j'}\over y_j\right)
\langle t(x_1)\ldots t(x_K)\rangle_a.
$$
Therefore
\eq{
\langle t(x_1)\ldots t(x_K)s(y_1)\ldots s(y_L)\rangle_a
=\langle t(x_1)\ldots t(x_K)s(y_1)\ldots s(y_L)\rangle_{-a}.
\label{tsinv}
}
Our aim is to prove the reflection property for all descendant form factors from this identity. The plan is as follows. First, we prove that products of the vertex operators $t(x_i)$ and $s(y_j)$ acting on the bra-vacuum ${}_a\langle1|$ span the whole right Fock module of the Heisenberg algebra~(\ref{ddcommut}). Similarly, these products acting on the ket-vacuum $|1\rangle_a$ span the whole left Fock module. Hence, the reflection map acts on the space of the matrix elements of the operator $t(x_1)\ldots t(x_N)$ with respect to the whole Fock modules. Next we impose the restriction to the states (\ref{hstatesdef}) and check that this restriction does not break the $a\to-a$ symmetry. This will prove the reflection property in the chiral sector.

The idea to express the form factors of descendant operators in terms of some asymptotics of the form factors of primary operators belongs to Fateev, Postnikov and Pugai~\cite{Fateev:2006js}. They applied it to get the form factors of some descendants in the case of ${\cal Z}_N$ Ising models. This proof gives a firm basis for this procedure in the case of sine/h-Gordon model and explains some heuristic tricks used in~\cite{Fateev:2006js}.

Consider the expansion
\eq{
{}_a\langle1|t(\xi_1^{-1}z)\ldots t(\xi_k^{-1}z)
s(\eta_1^{-1}z)\ldots s(\eta_l^{-1}z)
=\sum^\infty_{n=0}z^{-n}\cdot{}_a\langle n;\xi_1,\ldots,\xi_k;
\eta_1,\ldots,\eta_l|.
\label{ntsstates}
}
For shortness, we shall denote $\Xi=(\xi_1,\ldots,\xi_k)$, $\Eta=(\eta_1,\ldots,\eta_l)$.
Evidently,
\Align{
{}_a\langle0;\Xi;\Eta|
&=(2\cos\pi a)^k\cdot{}_a\langle1|,
\notag
\\
{}_a\langle1;\Xi;\Eta|
&={}_a\langle1|(\Sigma^-_1(\Xi;\Eta)d^-_1+\Sigma^+_1(\Xi;\Eta)d^+_1),
\notag
\\
{}_a\langle2;\Xi;\Eta|
&={}_a\langle1|(\Sigma^-_2(\Xi;\Eta)d^-_2+\Sigma^+_2(\Xi;\Eta)d^+_2
+\Sigma^{--}_{11}(\Xi;\Eta)(d^-_1)^2+\Sigma^{++}_{11}(\Xi;\Eta)(d^+_1)^2
+\Sigma^{-+}_{11}(\Xi;\Eta)d^-_1d^+_1)
\notag
}
with some functions $\Sigma^\pm_1$, $\Sigma^\pm_2$,\dots\ The conjecture is that at each given level $n$ for large enough values of $k$, $l$ it is possible to choose a set $(\Xi^{(i)};\Eta^{(i)})$, $i=1,\ldots,\dim(\cF^{\otimes2})_n$ so that the vectors ${}_a\langle n;\Xi^{(i)};\Eta^{(i)}|$ form a basis in the whole Fock module of the Heisenberg algebra~(\ref{ddcommut}).

First, prove this conjecture for the limiting case $a\to-\i\infty$, where $t(x)$ is proportional to $\lambda_-(x)$. Consider the product
\Multline{
\lambda_-(\xi_1^{-1}z)\ldots\lambda_-(\xi_k^{-1}z)
s(\eta_1^{-1}z)\ldots s(\eta_l^{-1}z)
\\
=\prod^k_{i=1}\prod^l_{j=1}f\left(\xi_i\over\eta_j\right)
\prod^l_{j<j'}f\left(\eta_j\over\eta_{j'}\right)f\left(\eta_{j'}\over\eta_j\right)
\lcolon\lambda_-(\xi_1^{-1}z)\ldots\lambda_-(\xi_k^{-1}z)
s(\eta_1^{-1}z)\ldots s(\eta_l^{-1}z)\rcolon.
\notag
}
The normal product in the right hand side is equal to
$$
\lcolon\lambda_-(\xi_1^{-1}z)\ldots\lambda_-(\xi_k^{-1}z)
s(\eta_1^{-1}z)\ldots s(\eta_l^{-1}z)\rcolon
=\lcolon\exp\left(\sum_{n\ne0}{\tau^-_nd^-_n+\tau^+_nd^+_n\over n}z^{-n}\right)\rcolon,
$$
where
$$
\tau^-_n=\sum^k_{i=1}\xi_i^n+\sum^l_{j=1}\eta_j^n,
\qquad
\tau^+_n=(-)^n\sum^l_{j=1}\eta_j^n.
$$
Consider the expansion
$$
{}_a\langle1|\lcolon\lambda_-(\xi_1^{-1}z)\ldots\lambda_-(\xi_k^{-1}z)
s(\eta_1^{-1}z)\ldots s(\eta_l^{-1}z)\rcolon
=\sum^\infty_{n=0}z^{-n}\cdot{}_{(-)}\langle n;\Xi;\Eta|.
$$
Then
$$
{}_{(-)}\langle n;\Xi;\Eta|
={}_a\langle1|\sum^n_{s=1}\sum_{n_1,\ldots,n_s>0\atop n_1+\cdots+n_s=n}
C_{n_1\ldots n_s}\prod^s_{i=1}(\tau^-_{n_i}d^-_{n_i}+\tau^+_{n_i}d^+_{n_i})
$$
with some nonzero coefficients~$C_{n_1\ldots n_s}$. It means that all possible products of $d^\pm_{n_i}$ enter the right hand side.

For large enough $k$, $l$ the functions $\tau^\pm_1,\ldots,\tau^\pm_n$ are functionally independent and they can be considered as independent variables. Besides, the monomials $\tau^{\ve_1}_{n_1}\dots\tau^{\ve_s}_{n_s}$ are linearly independent. Hence, for any set of the numbers $A^{\ve_1\ldots\ve_s}_{n_1\ldots n_s}$, $s=1,\ldots,n$, $n_1,\ldots,n_s>0$, $n_1+\cdots+n_s=n$, we have
$$
\sum_s\sum_{\ve_1,\ldots,\ve_n\atop n_1,\ldots,n_s}
\overline{A^{\ve_1\ldots\ve_s}_{n_1\ldots n_s}}
\>\tau^{\ve_1}_{n_1}\dots\tau^{\ve_s}_{n_s}\ne0
$$
for some values of the variables $\tau^\pm_1,\ldots,\tau^\pm_n$. Therefore, the vector generated by the numbers $A^{\ve_1\ldots\ve_s}_{n_1\ldots n_s}$ is not orthogonal to some vector generated by products of $\tau^\pm_m$. It means that there is no vector in the $\dim(\cF^{\otimes2})_n$-dimensional space orthogonal to all vectors generated by products of $\tau^\pm_m$ for any values of $\Xi$,~$\Eta$. It proves that the vectors of the form ${}_{(-)}\langle n;\Xi;\Eta|$ for some values $\Xi^{(i)}$, $\Eta^{(i)}$, $i=1,\ldots,\dim(\cF^{\otimes2})_n$ form a basis in the level $n$ subspace of the Fock module.

Since for generic values of $a$ the vertex operators $t(z)/2\cos\pi a$ are continuous deformations of the operator $\lambda_-(z)$, this proves that the vectors ${}_a\langle n;\Xi;\Eta|$ also span the level $n$ subspace of the right Fock module for generic values of~$a$. The proof for the left Fock module is just the same.

Let ${}_a\langle n;i|={}_a\langle n;\Xi^{(i)};\Eta^{(i)}|$, $i=1,\ldots,\dim(\cF^{\otimes2})_n$ be basic vectors in the level $n$ subspace of the right Fock module. Let $|\bar n;j\rangle_a$ be basic vectors in the level $\bar n$ subspace of the left Fock module. Then we have
$$
{}_a\langle n;i|t(x_1)\ldots t(x_N)|\bar n;j\rangle_a
={}_{-a}\langle n;i|t(x_1)\ldots t(x_N)|\bar n;j\rangle_{-a}.
$$
Now we have to select the vectors generated by $\pi_R(h)$, $\pi_L(h)$ for $h\in\cA$. The vector $\langle v|$ has the form ${}_a\langle1|\pi_R(h)$ for some $h$ if and only if
\subeq{\eq{
\langle v|(d^-_{-m}+(-1)^md^+_{-m})=0,\qquad m>0.
\label{leftred}
}
Similarly, the vector $|v\rangle$ has the form $\pi_L(h)|1\rangle_a$ if and only if
\eq{
(d^-_m+(-1)^md^+_m)|v\rangle=0,\qquad m>0.
\label{rightred}
}\label{redcond}}
Let the vectors ${}_a\langle1|\pi_R(h_{a,n,\mu})={}_a\langle\widetilde{n;\mu}|=\sum_iv^\mu_i(a)\>{}_a\langle n;i|$ form a basis in the $(\dim\cF_n)$-dimensional subspace of the level $n$ subspace of the Fock module defined by the conditions~(\ref{leftred}). Similarly define the vectors $\pi_L(h'_{a,\bar n,\nu})|1\rangle_a=|\widetilde{\bar n;\nu}\rangle_a=\sum_j\bar v^\nu_j(a)|\bar n;j\rangle_a$. It is easy to check from the commutation relations that
\Multline{
\langle t(x_1\ldots t(x_K)s(y_1)\ldots s(y_L)(d^-_{-m}+(-1)^md^+_{-m})
t(x'_1\ldots t(x'_{K'})s(y'_1)\ldots s(y'_{L'})\rangle_a
\\
=\langle t(x_1\ldots t(x_K)s(y_1)\ldots s(y_L)(d^-_{-m}+(-1)^md^+_{-m})
t(x'_1\ldots t(x'_{K'})s(y'_1)\ldots s(y'_{L'})\rangle_{-a}.
\notag
}
We have
\Align{
0
&={}_a\langle\widetilde{n;\mu}|(d^-_{-m}+(-1)^md^+_{-m})|n-m;j\rangle_a
\notag
\\
&=\sum_iv^\mu_i(a)\>{}_a\langle n;i|(d^-_{-m}+(-1)^md^+_{-m})|n-m;j\rangle_a
\notag
\\
&=\sum_iv^\mu_i(a)\>{}_{-a}\langle n;i|(d^-_{-m}+(-1)^md^+_{-m})|n-m;j\rangle_{-a}
\notag
}
Therefore,
$$
\sum_iv^\mu_i(a)\>{}_{-a}\langle n;i|(d^-_{-m}+(-1)^md^+_{-m})=0
$$
and there exists an element $\tilde h_{-a,n,\mu}$ such that
$$
\sum_iv^\mu_i(a)\>{}_{-a}\langle n;i|={}_{-a}\langle1|\pi_R(\tilde h_{-a,n,\mu}).
$$
Similarly, there exists an element $\tilde h'_{-a,\bar n,\nu}$ such that
$$
\sum_j\bar v^\nu_j(a)|\bar n;j\rangle_{-a}=\pi_L(\tilde h'_{-a,\bar n,\mu})|1\rangle_{-a}.
$$
Besides, we have
\Multline{
\langle\pi_R(h_{a,n,\mu})t(x_1)\ldots t(x_N)\pi_L(h'_{a,\bar n,\nu})\rangle_a
={}_a\langle\widetilde{n;\mu}|t(x_1)\ldots t(x_N)
|\widetilde{\bar n,\nu}\rangle_a
\\
=\sum_{i,j}v^\mu_i(a)\bar v^\nu_j(a)\>{}_a\langle n;i|t(x_1)\ldots t(x_N)|\bar n,j\rangle_a
=\sum_{i,j}v^\mu_i(a)\bar v^\nu_j(a)\>{}_{-a}\langle n;i|t(x_1)\ldots t(x_N)|\bar n,j\rangle_{-a}
\\
=\langle\pi_R(\tilde h_{-a,n,\mu})t(x_1)\ldots t(x_N)
\pi_L(\tilde h'_{-a,\bar n,\nu})\rangle_{-a}.
\notag
}
Comparing with eq.~(\ref{reflection}) we conclude that
$$
r_a(h_{a,n,\mu})=\tilde h_{-a,n,\mu},
\qquad
r_{-a}(h'_{a,\bar n,\nu})=\tilde h'_{-a,\bar n,\nu},
$$
which proves the reflection property.

\subsection*{Example: derivation of the second level reflection properties}

Now let us rederive the reflection properties (\ref{reflection(2,0)}) at the level $(2,0)$ using the construction described in this section. The advantage of this derivation is that it immediately proves the reflection relations for the level $(0,2)$ and $(2,2)$ cases.

To get rid of excessive functions $f(z)$ related to normal ordering, introduce the notation ${}_a^*\langle n;\Xi;\Eta|$:
\eq{
{}_a\langle n;\Xi;\Eta|
=\prod_{i=1}^k\prod_{j=1}^lf\left(\xi_i\over\eta_j\right)
\prod^l_{j<j'}f\left(\eta_j\over\eta_{j'}\right)f\left(\eta_{j'}\over\eta_j\right)
\cdot{}_a^*\langle n;\Xi;\Eta|
\label{ntsstarstates}
}

Let us search for the bra-vector in the form
\eq{
{}_a\langle X|=X_1\cdot{}_a^*\langle2;\xi_1,\xi_2|
+X_2\cdot{}_a^*\langle2;\xi;\eta|.
\label{Xform}
}
Let $\rho=\e^{\i\pi a}$. Then we have
\AlignS{
{}_a^*\langle0;\xi_1,\xi_2|
&={}_a\langle1|(\rho+\rho^{-1})^2,
\\
{}_a^*\langle0;\xi;\eta|
&={}_a\langle1|(\rho+\rho^{-1}),
\\
{}_a^*\langle1;\xi_1,\xi_2|
&={}_a\langle1|{1\over\xi_1+\xi_2}\Bigl(
((\rho^2+1)(\xi_1+\xi_2)^2-(\omega-\omega^{-1})\xi_1\xi_2)d^-_1
\\*
&\quad
+((\rho^{-2}+1)(\xi_1+\xi_2)^2+(\omega-\omega^{-1})\xi_1\xi_2)d^+_1
\Bigr),
\\
{}_a^*\langle1;\xi;\eta|
&={}_a\langle1|\Bigl((\rho\xi+(\rho+\rho^{-1})\eta)d^-_1
+(\rho^{-1}\xi-(\rho+\rho^{-1})\eta)d^+_1\Bigr),
\\
{}_a^*\langle2;\xi_1,\xi_2|
&={}_a\langle1|{1\over2}\Bigl(
((\rho^2+1)(\xi_1+\xi_2)^2-(\omega-\omega^{-1}+2\rho^2+2))\xi_1\xi_2)d^-_2
\\*
&\quad
+((\rho^{-2}+1)(\xi_1+\xi_2)^2+(\omega-\omega^{-1}-2\rho^{-2}-2))
\xi_1\xi_2)d^+_2
\\*
&\quad
+((\rho^2+1)(\xi_1+\xi_2)^2-(\omega-\omega^{-1}+2))(d^-_1)^2
\\*
&\quad
+((\rho^{-2}+1)(\xi_1+\xi_2)^2+(\omega-\omega^{-1}-2))(d^+_1)^2
+2\xi_1\xi_2d^-_1d^+_1
\Bigr),
\\
{}_a^*\langle2;\xi;\eta|
&={}_a\langle1|{1\over2}\Bigl(
(\rho\xi^2+(\rho+\rho^{-1})\eta^2)d^-_2
+(\rho^{-1}\xi^2+(\rho+\rho^{-1})\eta^2)d^+_2
\\*
&\quad
+(\rho\xi^2+(\rho+\rho^{-1})\eta^2+2\rho\xi\eta)(d^-_1)^2
+(\rho^{-1}\xi^2+(\rho+\rho^{-1})\eta^2-2\rho^{-1}\xi\eta)(d^-_1)^2
\\*
&\quad
-2\eta(\xi(\rho-\rho^{-1})+\eta(\rho+\rho^{-1}))d^-_1d^+_1
\Bigr).
}

From the definitions (\ref{ntsstates}), (\ref{ntsstarstates}) it is easy to derive that
\eq{
{}_a^*\langle n;\Xi;\Eta|{d^-_{-m}+(-)^md^+_{-m}\over A^+_m}
=(S_m(\Xi)+(1+(-1)^m)S_m(\Eta))\cdot
{}_a^*\langle n-m;\Xi;\Eta|.
\label{invaction}
}
Hence, the condition (\ref{leftred}) for the vector ${}_a\langle X|$ is equivalent to the system of homogeneous linear equations
$$
\pMatrix{(\rho+\rho^{-1})(\xi_1^2+\xi_2^2)&\xi^2+2\eta^2
\\
(\rho^2+1)(\xi_1+\xi_2)^2-(\omega-\omega^{-1})\xi_1\xi_2
&\xi(\rho\xi+(\rho+\rho^{-1})\eta)
\\
(\rho^{-2}+1)(\xi_1+\xi_2)^2+(\omega-\omega^{-1})\xi_1\xi_2
&\xi(\rho^{-1}\xi-(\rho+\rho^{-1})\eta)}
\pMatrix{X_1\\X_2}=0.
$$
The consistency condition to this system reduces to
$$
{(\xi_1+\xi_2)^2\over\xi_1\xi_2}
=-{(\omega-\omega^{-1})^2\over(\rho+\rho^{-1})^4},
\qquad
{\xi\over\eta}
={\omega-\omega^{-1}\over(\rho+\rho^{-1})^2},
$$
while its solution is
$$
{X_2\over X_1}=(\rho+\rho^{-1}){\xi_1\xi_2\over\eta^2}.
$$
Substituting this to (\ref{Xform}) we get
$$
\Aligned{
{}_a\langle X|
&={}_a\langle1|(-\i)(\sin\pi p+\sin2\pi a)
(d^-_2-d^+_2-\i\tg\pi a\>(d^-_1-d^+_1)^2)
\\
&\quad
+{}_a\langle1|(2\cos2\pi a-1)(d^-_1-d^+_1)^2.
}
$$
The first term is proportional to the $(\sin^2\pi p-\sin^22\pi a)\cdot{}_a\langle1|\pi_R(h^{(2)}_a)$, while the second is proportional to $(2\cos2\pi a-1)\cdot{}_a\langle1|\pi_R(h^{(1,1)}_a)$. Due to the evident $a\to-a$ invariance of~$h^{(1,1)}_a$,%
\footnote{In the notation used in this section ${}_a\langle1|\pi_R(h^{(1,1)}_a)={}_a^*\langle2;;\i^{1/2},\i^{-1/2}|$.}
it proves the invariance of~$h^{(2)}_a$.

{\bf Remark.} Such kind of derivation can be essentially simplified by use of the `even' projectors:
\eq{
\Gathered{
P_{2k}=\lcolon\exp\left(
-{(d^-_{-2k}+d^+_{-2k})(d^-_{2k}+d^+_{2k})\over4kA^+_{2k}}
\right)\rcolon,
\qquad
k>0;
\\
P_{2k}^2=P_{2k},
\qquad
[P_{2k},P_{2l}]=0,
\qquad
P_{2k}(d^-_{-2k}+d^+_{-2k})=(d^-_{2k}+d^+_{2k})P_{2k}=0.
}\label{enenprojectors}
}
These projectors are $a\to-a$ invariant, i.~e.\ insertion of such projectors into any (but the same) places at both sides of (\ref{tsinv}) does not break the identity. The vector ${}_a\langle n;\Xi;\Eta|P_2P_4\ldots$ automatically satisfies the equation (\ref{leftred}) with even values of~$m$. This reduces the number of equations to be solved. Unfortunately, `odd' projectors cannot be made two-sided nor $a\to-a$ invariant.

\section{The kink sector}

Now let us propose a conjecture about the kink sector.

Let $V(\theta)$ and $\bar V(\gamma)$ be the vertex operators defined as~\cite{Lukyanov:1997bp}
\eq{
\Aligned{
\llangle V(\theta_2)V(\theta_1)\rrangle
&=G(\theta_1-\theta_2),
\\
\llangle V(\theta_2)\bar V(\theta_1)\rrangle
=\llangle V(\theta_2)\bar V(\theta_1)\rrangle
&=W(\theta_1-\theta_2)
\equiv G^{-1}(\theta_1-\theta_2-\i\pi/2)G^{-1}(\theta_1-\theta_2+\i\pi/2),
\\
\llangle\bar V(\theta_2)\bar V(\theta_1)\rrangle
&=\bar G(\theta_1-\theta_2)
\equiv W^{-1}(\theta_1-\theta_2-\i\pi/2)W^{-1}(\theta_1-\theta_2+\i\pi/2)
}\label{VVdef}
}
with
\Align{
G(\theta)&=\exp\left(-\int^\infty_0{dt\over t}\,
{\sh{\pi t\over2}\sh{\pi(p+1)t\over2}\ch(\pi-\i\theta)t
\over\sh^2\pi t\sh{\pi pt\over2}}\right)
\notag
\\
&=\i{\e^{C_E}\over\pi}\sh{\theta\over\pi}
\exp\int^\infty_0{dt\over t}\,
{\sh{\pi t\over2}\sh{\pi(p-1)t\over2}\ch(\pi-\i\theta)t
\over\sh^2\pi t\sh{\pi pt\over2}}.
\label{Gdef}
}
Here $C_E$ is the Euler constant. Formally the integrals diverge at zero, and to make them convergent we define them as follows:
$$
\int^\infty_0dt\,f(t)=\left.\int^\infty_\epsilon dt\,f(t)-{1\over\epsilon}\Res_{t=0}tf(t)
+\log\epsilon\cdot\Res_{t=0}f(t)\right|_{\epsilon\to0},
$$
if $f(t)$ possesses a double pole at zero.

Let
\Align{
V(\theta)
&=V(\theta)\,\e^{{\hat a+1/2\over p}\theta},
\label{Vthetadef}
\\
S(\theta)
&=\int_{\cC(\theta)}{d\gamma\over2\pi}\,\bar V(\theta)\,
{\e^{-2{\hat a+1/2\over p}\gamma}
\over\sh{\gamma-\theta-\i\pi/2\over p}}.
\label{Sdef}
}
As usual the contour $\cC(\theta)$ goes along the real axis with a fold: it goes above $\theta+\i\pi/2$ and below $\theta-\i\pi/2$.

Then let
\eq{
\Aligned{
Z_+(\theta)
&=\lambda V(\theta),
\\
Z_-(\theta)
&=\i\lambda\bar\lambda V(\theta)S(\theta).
}\label{Zpmdef}
}
Here
$$
\Aligned{
\lambda
&=\exp\left(-\int^\infty_0{dt\over t}\,
{\e^{-\pi t}\sh{\pi t\over2}\sh{\pi(p+1)t\over2}
\over2\sh^2\pi t\sh{\pi pt\over2}}\right)
\\
\bar\lambda
&={\e^{2{p+1\over p}(C_E+\log\pi p)}\over\pi p^4}
{\Gamma\left(1\over p\right)\over\Gamma\left(1-{1\over p}\right)}
\exp\left(-\int^\infty_0{dt\over t}\,{\e^{-3\pi t/2}\sh{\pi(p+1)t\over2}
\over\sh\pi t\sh{\pi pt\over2}}\right).
}
$$

The form factors of primary operators are given by
\eq{
G_af_a(\theta_1,\ldots,\theta_N)_{\ve_1\ldots\ve_N}
=G_a\llangle Z(\theta_N)\ldots Z(\theta_1)\rrangle_a.
\label{falphakinkdef}
}
Now we define a generalization of these form factors.

Let $\ve_1,\ldots,\ve_N=\pm$. Let us define the numbers
\eq{
\{s_j\}^M_{j=1}=\{i\>|\>\ve_i=-\},
\qquad
s_1<s_2<\ldots<s_M,
\label{ssetdef}
}
Let
\eq{
V_i=\Cases{V(\theta_i),&\epsilon_i=+,
\\
V(\theta_i)\bar V(\gamma_j),&\epsilon_i=-,\quad i=s_j.}
\label{Videf}
}
Let us search the form factors in the form
\eq{
\Split{
f_a^Q(\theta_1,\ldots,\theta_N)_{\ve_1\ldots\ve_N}
&=(\i\eta)^{-M}\prod^M_{j=1}\int_{\cC(\theta_{s_j})}{d\gamma_j\over2\pi}\,
{1\over\sh{\gamma_j-\theta_{s_j}-\i\pi/2\over p}}\,
\llangle V_N\ldots V_1\rrangle
\\
&\times
\e^{{\alpha\over\beta}\left(\sum^N_{i=1}\theta_i
-2\sum^M_{j=1}\gamma_j\right)}
\\
&\times
Q_{N,M}(\e^{\theta_1},\ldots,\e^{\theta_N}|\e^{\gamma_1},\ldots,\e^{\gamma_M}).
}\label{fkink}
}
For $Q_{N,M}=1$ this is just the explicit form of~(\ref{falphakinkdef}).

The functions $Q_{N,M}(x_1,\ldots,x_N|z_1,\ldots,z_M)$ are rational subject to three conditions:

1. They are symmetric with respect to $\{x_i\}^N_{i=1}$ and $\{z_j\}^M_{j=1}$ separately:
\eq{
Q_{N,M}(\sigma X|Z)=Q_{N,M}(X|\tau Z)=Q_{N,M}(X|Z)
\label{Qk-sym}
}
for any permutations $\sigma\in S_N$ and $\tau\in S_M$.

2. They satisfy the chain equation
\eq{
Q_{N+2,M+1}(X,x,-x|Z,\i x)=Q_{N,M}(X|Z).
\label{Qk-kin}
}

3. They admit factorization property
$$
Q_{N,M}(X|Z)=\sum_AQ^A_{N,M}(X|Z)\bar Q^A_{N,M}(X^{-1}|Z^{-1})
$$
with the polynomials $Q^A_{N,M}$, $\bar Q^A_{N,M}$ being of the form~$P^{[p]}_{N+M,(N-M)/2}(X|Z)$ from~(\ref{Pk-p}).

We do not impose any restriction on the growth of $Q_{N,M}$ as $\gamma_j\to\pm\infty$. Since $G(\theta)\sim\e^{-{p+1\over2p}|\theta|}$ as $\theta\to\pm\infty$, if $Q$ is a polynomial in $z_i$, $z_i^{-1}$ the integrations in (\ref{fkink}) are convergent for small enough values of the parameter~$p$. The conjecture is that the integrals can be defined by an analytic continuation in~$p$.

The breather functions $P$ are related to the kink functions $Q$ as
\eq{
P(X_-|X_+)=Q_{2N,N}(-\i\omega^{1/2}X,\i\omega^{-1/2}X|
\omega^{-1/2}X_-,\omega^{1/2}X_+),
\qquad
X=X_-\cup X_+.
\label{PQrel}
}

First let the functions $Q_{N,M}$ be polynomials. We conjecture that physically it is equivalent to restriction to the right ($\cL_{-k}$) chirality. Let $\cQ_n$ be the space of the homogeneous polynomials of the order~$n$ subject to the conditions~1--3.

\begin{theorem}
The dimensions of the spaces $\cQ_n$ are given by the same generating function~$\chi(q)$.
\end{theorem}

The proof is quite similar and uses the same commutative algebra. Let
\eq{
K_n=2\i^{1-n}\sin{\pi pn\over2}.
\label{Kmdef}
}
Let us slightly change the basic elements of the algebra $\cA$:
\eq{
C_{-m}=K_m^{-1}c_{-m}.
\label{Cmdef}
}
Let
\eq{
A(z)=\e^{\sum^\infty_{m=1}C_{-m}z^m},
\qquad
D(z)=\e^{2\sum^\infty_{m=1}(-1)^{m-1}C_{-2m}z^{2m}}.
\label{dcurrent}
}
Then
\Align{
A(x)A(-x)D(\i x)
&=1,
\label{AADprod}
\\
A(-\i\omega^{1/2}x)A(\i\omega^{-1/2}x)D(\omega^{\ve/2}x)
&=\Cases{a(x),\quad\ve=-1,\\b(x),\quad\ve=+1.}
\label{AADprodab}
}
Define the functions
\eq{
Q^h_{N,M}(X,Z)=(A(x_1)\ldots A(x_N)D(z_1)\ldots D(z_M),h).
\label{Qhdef}
}
These functions solve the equation~(\ref{Qk-kin}). The property (\ref{AADprodab}) provides the relation~(\ref{PQrel}). The proof of linear independence of the solutions corresponding to the elements of the form~(\ref{basicel}) is similar to that for the functions~$P^h_{N,k}(X|Y)$. The explicit form of these solutions is
\eq{
Q^h_{N,M}(X,Z)
={1\over K_h}\prod^{\lceil s/2\rceil}_{m=1}S_{2m-1}^{k_{2m-1}}(X)
\prod^{\lfloor s/2\rfloor}_{m=1}
(S_{2m}(X)+2(-1)^{m-1}S_{2m}(Z))^{k_{2m}},
\label{Qh-explicit}
}
where
\eq{
K_h=\prod^{\lceil s/2\rceil}_{m=1}K_{2m-1}^{k_{2m-1}}
\prod^{\lfloor s/2\rfloor}_{m=1}K_{2m}^{k_{2m}}.
\label{Khdef}
}
For the `antichiral' algebra $\bar\cA$ we similarly define
\eq{
\bar A(z)=\e^{\sum^\infty_{m=1}\bar C_{-m}z^m},
\qquad
\bar D(z)=\e^{2\sum^\infty_{m=1}(-1)^{m-1}\bar C_{-2m}z^{2m}}.
\label{bardcurrent}
}

Now we can write down the fully algebraic representation in the kink sector. In similar notation as (\ref{fhdef}) define the bare vertex and the screening operator
\Align{
\cV(\theta)
&=A(\e^\theta)\bar A(\e^{-\theta})V(\theta)
\e^{{\hat a+1/2\over p}\theta},
\label{cVdef}
\\
\cS(\theta)
&=\int_{\cC(\theta)}{d\gamma\over2\pi}\,
D(\e^{\gamma})\bar D(\e^{-\gamma})\bar V(\gamma)
{\e^{-2{\hat a+1/2\over p}\gamma}\over\sh{\gamma-\theta-\i\pi/2\over p}}.
\label{cSdef}
}
Let
\eq{
\Aligned{
\cZ_+(\theta)
&=\lambda\cV(\theta),
\\
\cZ_-(\theta)
&=\i\lambda\bar\lambda\cV(\theta)\cS(\theta).
}\label{cZpmdef}
}
Let $g\in\cA\otimes\bar\cA$. Then the kink form factors of the operator $V_a^g(x)$ read
\eq{
G_af^g_a(\theta_1,\ldots,\theta_N)_{\ve_1\ldots\ve_N}
=G_a(\llangle\cZ_{\ve_N}(\theta_N)\ldots
\cZ_{\ve_1}(\theta_1)\rrangle_a,g).
\label{fhkinkdef}
}
The field $\tilde V^g_a$ is still defined by~(\ref{VtildeVrel}).

\section{Conclusion}

The results of the present work extend the applicability of the free field representation to descendant operators. The algebraic receipt presented here seems to admit rather straightforward generalization to other theories. The auxiliary free field representation is likely to be more specific, but it is clear that it also can be generalized to, for example, affine Toda theories. Hopefully, the proof of the existence of reflection relations can be also generalized to this case. Another way to develop the results presented here is to study truncations of the spaces of operators at rational values of~$p$. Though there are many important results on counting descendant operators in the restricted sine-Gordon theory (see e.~g.\ \cite{Babelon:1996sk,Jimbo:2003ge,Delfino:2007bt}), it is important to clarify the restriction procedure from the point of view of the free field approach.

The most important and ambitious problem that probably can be addressed with the help of these results is to find a way to identify the form factors at each level with the particular descendant operators obtained from the exponential ones by means of the Heisenberg algebra (\ref{varphi-expansion}), (\ref{PQaa-alg}) or of the Virasoro algebra.

\section{Acknowledgments}

We are grateful to P.~Baseilhac, M.~Jimbo, Ya.~Pugai, S.~Roan, J.~Shiraishi, F.~Smirnov, and A.~Zamolodchikov for interesting and stimulating discussion. The work was, in part, supported by the Russian Foundation of Basic Research (the grants 08--01--00720, 05--01--02934) and by the Program for Support of Leading Scientific Schools (the grant 3472.2008.2). Besides, the visit of M.~L.\ to LMPT, Universit\'e de Tours in October 2007 and LPTHE, Universit\'e Paris 6 in August--September 2008 was supported by the ENS--Landau Exchange Program.

\appendix
\makeatletter
\def\@seccntformat#1{Appendix.~~}
\makeatother
\section{Equation of Motion and Energy-Momentum Conservation}

\makeatletter
\def\@seccntformat#1{\csname the#1\endcsname.~~}
\makeatother
\subsection{Equation of Motion}

Our aim is to prove that the form factors are consistent with the equation of motion
\eq{
\d\bar\d\varphi=\pi\mu\beta\sin\beta\varphi.
\label{eqmotion}
}
Though this fact has already been proven in~\cite{Babujian:2002fi}, it is instructive to rederive it from the recursion relation~(\ref{gNNrec}).

The derivatives of a field produce multiplication of its form factors by the components of the momentum according to the usual rule $P_\mu\leftrightarrow\i\d_\mu$. These components are given by
$$
\Aligned{
P_z(\theta_1,\ldots,\theta_N)
&=-{m\over2}\sum^N_{i=1}\e^{\theta_i}=-{m\over2}S_1(X)=-{m\over2}P^{c_{-1}}(X_-|X_+),
\\
P_{\bz}(\theta_1,\ldots,\theta_N)
&={m\over2}\sum^N_{i=1}\e^{-\theta_i}={m\over2}S_{-1}(X)={m\over2}P^{\bar c_{-1}}(X_-|X_+).
}
$$
Since $\varphi(x)=-\i\left.d\e^{\i\alpha\varphi}/d\alpha\right|_{\alpha=0}$ and $\left.d\langle\e^{\i\alpha\varphi}\rangle/d\alpha\right|_{\alpha=0}=0$, we have
$$
\langle0|\d\bar\d\varphi(0)|\theta_1,\ldots,\theta_N\rangle
={m^2\over4\i}\sqrt{p(p+1)\over2}
\left(\sum^N_{i=1}\e^{\theta_i}\right)\left(\sum^N_{i=1}\e^{-\theta_i}\right)
\left.{d\over da}f_a(\theta_1,\ldots,\theta_N)\right|_{a=-1/2}.
$$
Due to the reflection property (\ref{exp-reflection}) these form factors vanish for even values of~$N$. On the other hand, since~\cite{Lukyanov:1996jj}
\eq{
\mu\langle\e^{\i\beta\varphi}\rangle=m^2{1+p\over8\sin\pi p},
\label{muGprod}
}
we have
$$
\Aligned{
\pi\mu\beta\langle0|\sin\beta\varphi|\theta_1,\ldots,\theta_N\rangle
&={\pi\over2\sin\pi p}{m^2\over4\i}\sqrt{p(p+1)\over2}\>
(f_{p-1/2}(\theta_1,\ldots,\theta_N)-f_{-p-1/2}(\theta_1,\ldots,\theta_N))
\\
&=\Cases{0,&N\in2\Z,\\
{\pi\over\sin\pi p}{m^2\over4\i}\sqrt{p(p+1)\over2}
f_{p-1/2}(\theta_1,\ldots,\theta_N),&N\in2\Z+1.}
}
$$
The last equality follows from (\ref{exp-reflection}),~(\ref{fperiodicity}).

Let
\eq{
J'_N(x)=\left.{d\over da}J_{N,a}(X)\right|_{a=-1/2},
\qquad
R'_{N,i}(x)=\left.{d\over da}R_{N,a,i}(X)\right|_{a=-1/2}.
\label{J'R'def}
}
Then the equation of motion can be rewritten as
\eq{
S_1(X)S_{-1}(X)J'_N(X)={\pi\over\sin\pi p}J_{N,p-1/2}(X)
\qquad
\text{for odd $N$.}
\label{Jeqmotion}
}
Let us prove this identity by induction. For the function in the right hand side the recursion relation (\ref{gNNrec}) takes the form
\eq{
J_{N,p-1/2}(X,x)=2\sin\pi p\>J_{N-1,p-1/2}(X)
+\sum^{N-1}_{i=1}{x_i\over x+x_i}R_{N,p-1/2,i}(X).
\label{Jp-1/2rec}
}
For the derivative in the left hand side we have
\eq{
\Aligned{
J'_1(x)
&=2\pi,
\\
J'_N(X,x)
&=\sum^{N-1}_{i=1}{x_iR'_{N,i}(X)\over x+x_i}
\qquad(N=3,5,\ldots).
}\label{DJrec}
}
Rewrite the last line as
\eq{
S_1(X,x)S_{-1}(X,x)J'_1(X,x)
=\sum^{N-1}_{i=1}x_iS_{-1}(\hat X_i)R'_{N,i}(X)
+\sum^{N-1}_{i=1}S_1(\hat X_i)S_{-1}(\hat X_i){x_iR'_{N,i}(X)\over x+x_i}.
\label{DJrecmod}
}
Here we used the identity~(\ref{sumRprim}).

Now we want to use induction. The equation (\ref{Jeqmotion}) is evidently valid for $N=1$. Now suppose that it is valid for some odd value of $N$, which will be denoted from now on as $M-2$. Let us prove it for $N=M$. By the hypothesis of the induction we have
$$
S_1(\hat X_i)S_{-1}(\hat X_i)R'_{M,i}(X)
={\pi\over\sin\pi p}R_{M,p-1/2,i}(X).
$$
Hence, the first term in the right hand side of Eq.~(\ref{DJrecmod}) is equal to
$$
\Aligned{
{\pi\over\sin\pi p}\sum^{M-1}_{i=1}{x_i\over S_1(\hat X_i)}R_{M,p-1/2,i}(X)
&=-{\pi\over\sin\pi p}
\sum^{M-1}_{i=1}{x_i\over-S_1(X)+x_i}R_{M,p-1/2,i}(X)
\\*
&={\pi\over\sin\pi p}(2\sin\pi p\>J_{M-1,p-1/2}(X)
-J_{M,p-1/2}(X,-S_1(X)),
}
$$
while the second term is, according to the recurrent relation, equal to
$$
{\pi\over\sin\pi p}
\sum^{M-1}_{i=1}{x_i\over x+x_i}R_{M,p-1/2,i}(X)
={\pi\over\sin\pi p}(J_{M,p-1/2}(X,x)-2\sin\pi p\>J_{M-1,p-1/2}(X)).
$$
Collecting both terms yields
$$
S_1(X,x)S_{-1}(X,x)J'_N(X,x)
={\pi\over\sin\pi p}(J_{M,p-1/2}(X,x)-J_{M,p-1/2}(X,-S_1(X))).
$$
The function $J_{M,p-1/2}(X,-S_1(X))$ is $x$-independent. Since we can take for $x$ any element of the set $\{x_1,\ldots,x_M\}$, this function must be constant in all variables $x_1,\ldots,x_{M-1}$. From the recurrent equation we have
$$
\const=J_{M,p-1/2}(X,-S_1(X))
=2\sin\pi p\>J_{M-1,p-1/2}(X)+\sum^{M-1}_{i=1}{x_i\over S_1(\hat X_i)}
R_{M,p-1/2,i}(X).
$$
Since the left hand side is a constant, we may calculate it in the limit $x_{M-1}\to\infty$. In this limit the only nonvanishing term in the sum is that with $i=M-1$. We have
\Multline{
J_{M,p-1/2}(X,-S_1(X))
=(2\sin\pi p)^2J_{M-2,p-1/2}(\hat X_{M-1})
\\*
-\left[\i\sin\pi p\>{x_{M-1}\over S_1(\hat X_{M-1})}
\left(\prod_{j=1}^{M-2}f\left(x_{M-1}\over x_j\right)
-\prod_{j=1}^{M-2}f\left(x_j\over x_{M-1}\right)\right)
J_{M-2,p-1/2}(\hat X_{M-1})\right]_{x_{M-1}\to\infty}.
\notag
}
Since $f(x)=1+{2\i\sin\pi p\over x}$ as $x\to\infty$, the second term cancels the first one and we obtain
\eq{
J_{M,p-1/2}(X,-S_1(X))=0.
\label{JXsigmaX0}
}
This proves (\ref{Jeqmotion}) for $N=M$ and, hence, for any odd~$N$.

\subsection{Energy-Momentum Conservation Law}

The energy-momentum conservation law looks like
\eq{
\Aligned{
\bd T=\d\Theta,
\\
\d\bT=\bd\Theta.
}\label{em-conserv}
}
Here
\subeq{\Align{
T(z,\bz)
&=-2\pi T^{\text{Mink}}_{zz}(z,\bz)
=-{1\over2}(\d\varphi(z,\bz))^2,
\label{T-def}
\\
\bT(z,\bz)
&=-2\pi T^{\text{Mink}}_{\bz\bz}(z,\bz)
=-{1\over2}(\bd\varphi(z,\bz))^2,
\label{Tb-def}
\\
\Theta(z,\bz)
&=2\pi T^{\text{Mink}}_{z\bz}(z,\bz)
={\pi\mu\over1+p}\cos\beta\varphi(z,\bz).
\label{Theta-def}
}\label{Tcomp-def}}
The denominator $1+p$ in the second line is the well-known quantum correction to the potential part of the energy-momentum tensor in the sine-Gordon model.

The component $\Theta(x)$ is a combination of exponential fields, but the component $T(x)$ is a descendant and has to be identified. The last must be a linear combination of the operators $V_{-1/2}^{c_{-2}}(x)$ and $V_{-1/2}^{c_{-1}^2}(x)$. Let us prove that
\Align{
T(x)
&={\i\pi m^2\sin\pi p\over8}\left.V_a^{h^{(2)}_a}(x)\right|_{a\to-1/2},
\label{Tident}
\\
\bT(x)
&={\i\pi m^2\sin\pi p\over8}\left.V_a^{\bar h^{(2)}_{-a}}(x)\right|_{a\to-1/2},
\label{Tbident}
}
The $J$ functions corresponding to $V_a^{h^{(2)}_a}(x)$ are given by
$$
J^{h^{(2)}_a}_{N,a}(X)\simeq{1\over\sin\pi p}
\left(J^{c_{-2}}_{N,a}(X)+{\i\over\pi(a+1/2)}S_1^2(X)J_{N,a}(X)\right)
\quad
\text{as $a\to-1/2$.}
$$
The second term turns out to be finite in this limit since $J_{N,a}(x)\to0$ for $N>0$ and $S_1(X)=0$ for $N=0$. Hence,
\eq{
\left.J^{h^{(2)}_a}_{N,a}(X)\right|_{a\to-1/2}
={1\over\sin\pi p}
\left(J^{c_{-2}}_{N,-1/2}(X)+{\i\over\pi}S_1^2(X)J'_N(X)\right).
\label{Jh2-1/2def}
}

If (\ref{Tident}) is true, due to (\ref{Theta-def}) and (\ref{muGprod}) the first of the equations (\ref{em-conserv}) takes the form
\eq{
\left.J^{h^{(2)}_a\bar c_{-1}}_{N,a}(X)\right|_{a\to-1/2}
={\i\over2\sin^2\pi p}\left(J^{c_{-1}}_{N,-1/2+p}(X)+J^{c_{-1}}_{N,-1/2-p}(X)\right).
\label{em-conserv-J}
}
We want to prove this identity. First, notice that both the left and the right hand side of (\ref{em-conserv-J}) are zero for odd~$N$. Indeed, this is an immediate consequence of the reflection property together with the periodicity~(\ref{fperiodicity}). Hence, we have to prove the identity for even~$N$:
$$
\left.J^{h^{(2)}_a\bar c_{-1}}_{N,a}(X)\right|_{a\to-1/2}
={\i\over\sin^2\pi p}J^{c_{-1}}_{N,-1/2+p}(X),
\qquad
N\in2\Z.
$$
This identity is evidently true for $N=0$. Suppose that it is true for $N\le M-2$ for some value~$M$. According to the recurrent relation (\ref{J0form}) for $h^{(2)}_a$ with $J^{(0)}$ from~(\ref{J(0)h2}) we have
\Multline{
\left.J^{h^{(2)}_a\bar c_{-1}}_{M,a}(X,x)\right|_{a\to-1/2}
=-\sum^{M-1}_{i=1}
{x_i^{-1}\over x^{-1}+x_i^{-1}}\left.R^{h^{(2)}_a\bar c_{-1}}_{M,a,i}(X)\right|_{a\to-1/2}
-\sum^{M-1}_{i=1}\left.x_i^{-1}R^{h^{(2)}_a}_{M,a,i}(X)\right|_{a\to-1/2}
\\
+{2\i x\over\pi}(x^{-1}+S_{-1}(X))S_1(X)J'_{M-1}(X).
\notag
}
Applying the induction hypothesis to $R'_{M-2,i}$ and the equation (\ref{Jeqmotion}) to $J'_{M-1}$ we obtain
\Multline{
\left.J^{h^{(2)}_a\bar c_{-1}}_{M,a}(X,x)\right|_{a\to-1/2}
={\i\over\sin^2\pi p}\Biggl(
-\sum^{M-1}_{i=1}{x_i^{-1}R^{c_{-1}}_{M,p-1/2,i}(X)\over x^{-1}+x_i^{-1}}
\\
-\sum^{M-1}_{i=1}x_i^{-1}{S_1(X)-x_i\over S_{-1}(X)-x_i^{-1}}R_{M,p-1/2,i}(X)\Biggr)
+{2\i x\over\sin\pi p}{x^{-1}+S_{-1}(X)\over S_{-1}(X)}J_{M-1,p-1/2}(X).
\notag
}
On the other hand, from the recurrent relation (\ref{gNNrec}) and the identity (\ref{sumRprim}) we get
$$
J^{c_{-1}}_{M,p-1/2}(X,x)
=-\sum^{M-1}_{i=1}{x_i^{-1}R^{c_{-1}}_{M,p-1/2,i}(X)\over x^{-1}+x_i^{-1}}
+2\sin\pi p\>(x+S_1(X))J_{M-1,p-1/2}(X).
$$
Combining the last two equations we obtain
\Align{
&\left.J^{h^{(2)}_a\bar c_{-1}}_{M,a}(X,x)\right|_{a\to-1/2}
-{\i\over\sin^2\pi p}J^{c_{-1}}_{M,p-1/2}(X,x)
\notag
\\
&\qquad={\i\over\sin^2\pi p}\Biggl(
-\sum^{M-1}_{i=1}x_i^{-1}{S_1(X)-x_i\over S_{-1}(X)-x_i^{-1}}R_{M,p-1/2,i}(X)
\notag
\\
&\qquad\quad+2\sin\pi p\>\left(S_{-1}^{-1}(X)-S_1(X)\right)J_{M-1,p-1/2}(X)\Biggr)
\notag
\\
&\qquad={\i\left(S_{-1}^{-1}(X)-S_1(X)\right)\over\sin^2\pi p}\left(
-\sum^{M-1}_{i=1}x_i^{-1}{R_{M,p-1/2,i}(X)\over-S_{-1}(X)+x_i^{-1}}
+2\sin\pi p\>J_{M-1,p-1/2}(X)\right)
\notag
\\
&\qquad={\i\left(S_1(X)-S_{-1}^{-1}(X)\right)\over\sin^2\pi p}
J_{M,p-1/2}(X,-S_{-1}^{-1}(X))=0.
\notag
}
The last equality is derived in the same way as~(\ref{JXsigmaX0}). This proves~(\ref{em-conserv-J}) and, therefore,~(\ref{Tident}). The proof of~(\ref{Tbident}) is just the same.

We ought to make one more remark on the energy-momentum conservation law. Let us introduce two modified energy-momentum tensors $T^\pm_{\mu\nu}$ corresponding to the currents
\subeq{\Align{
T^\pm(z,\bz)
&=-{1\over2}(\d\varphi(z,\bz))^2\pm\i\alpha_0\d^2\varphi(x),
\label{Tpm-def}
\\
\bT^\pm(z,\bz)
&=-{1\over2}(\bd\varphi(z,\bz))^2\pm\i\alpha_0\bd^2\varphi(x),
\label{Tbpm-def}
\\
\Theta^\pm(z,\bz)
&={\pi\mu\over1+p}\e^{\pm\i\beta\varphi(z,\bz)}.
\label{Thetapm-def}
}\label{Tpmcomp-def}}
Due to the equation of motion they also satisfy the conservation laws
\eq{
\Aligned{
\bd T^\pm=\d\Theta^\pm,
\\
\d\bT^\pm=\bd\Theta^\pm.
}\label{empm-conservation}
}
These modified energy-momentum tensors are relevant to two kinds of quantum reduction in the sine-Gordon model. In the conformal limit the components $T^\pm$, $\bT^\pm$ generate the Virasoro algebras with the central charge $c=1-6/p(p+1)<1$.

Now, it is easy to check that
\Align{
&\Aligned{
T^+(x)
&={\i\pi m^2\over8}\left(V_{-1/2}^{(2)}(x)\sin\pi p
+{\i\over\pi}V_{-1/2}^{\prime c_{-1}^2}(x)\right)
={\i\pi m^2\over8}\left(V_{-1/2}^{c_{-2}}(x)
+{2\i\over\pi}V_{-1/2}^{\prime c_{-1}^2}(x)\right),
\\
\bT^+(x)
&={\i\pi m^2\over8}\left(V_{-1/2}^{(\bar2)}(x)\sin\pi p
+{\i\over\pi}V_{-1/2}^{\prime\bar c_{-1}^2}(x)\right)
={\i\pi m^2\over8}V_{-1/2}^{\bar c_{-2}}(x).
}\label{Tpc2}
\\
&\Aligned{
T^-(x)
&={\i\pi m^2\over8}\left(V_{-1/2}^{(2)}(x)\sin\pi p
-{\i\over\pi}V_{-1/2}^{\prime c_{-1}^2}(x)\right)
={\i\pi m^2\over8}V_{-1/2}^{c_{-2}}(x),
\\
\bT^-(x)
&={\i\pi m^2\over8}\left(V_{-1/2}^{(\bar2)}(x)\sin\pi p
-{\i\over\pi}V_{-1/2}^{\prime\bar c_{-1}^2}(x)\right)
={\i\pi m^2\over8}\left(V_{-1/2}^{\bar c_{-2}}(x)
-{2\i\over\pi}V_{-1/2}^{\prime\bar c_{-1}^2}(x)\right),
}\label{Tmc2}
}
where
$$
V_{-1/2}^{(2)}
=\left.V_a^{h^{(2)}_a}\right|_{a\to-{1\over2}},
\quad
V_{-1/2}^{(\bar2)}
=\left.V_a^{\bar h^{(2)}_{-a}}\right|_{a\to-{1\over2}},
\quad
V_a^{\prime g}(x)
=\left.{d\over da'}V_{a'}^g\right|_{a'\to a}.
$$

\end{document}